\documentclass[pre,preprint,12pt]{revtex4-1}
\usepackage[utf8]{inputenc}
\usepackage{graphicx,hyperref,subfig}

\usepackage{amsmath,amsfonts}

\newcommand{\PT}{$\mathcal{PT}$}

\begin{document}

\title{Statistical properties of eigenvalues of an ensemble of pseudo-Hermitian Gaussian matrices}

\author{G. Marinello}
\affiliation{Instituto de Física de São Carlos, São Carlos, Universidade de São Paulo}
\affiliation{Instituto de Física, Universidade de São Paulo}

\author{M. P. Pato}
\affiliation{Instituto de Física, Universidade de São Paulo}

\begin{abstract}
We investigate the statistical properties of eigenvalues of pseudo-Hermitian random matrices whose eigenvalues are real or complex conjugate. It is shown that when the spectrum splits into separated sets of real and complex conjugate eigenvalues, the real ones show characteristics of an intermediate incomplete spectrum, that is, of a so-called thinned ensemble. On the other hand, the complex ones show repulsion compatible with cubic-order repulsion of non normal matrices for the real matrices, but higher order repulsion for the complex and quaternion matrices.
\end{abstract}

\maketitle

\section{Introduction}

It can be shown that a complex non-Hermitian Hamiltonian invariant under the combined parity ($\mathcal{P}$) and time reversal ($\mathcal{T}$) transformations have eigenvalues which are real or complex conjugate. A Hamiltonian with this so-called \PT-symmetry is, for instance, 
\begin{equation}
	H=p^2 -(ix)^{\gamma}  
	\label{1}
\end{equation}
whose properties have been analyzed in a seminal paper \cite{Bender1998}. It was found that, as a function of the parameter $\gamma,$ for $\gamma > 2,$ eigenvalues are real and, progressively, as $\gamma$ decreases they move into the complex plane in conjugate pairs. This can be seen as a phase transition in which the system goes from a \PT-symmetric phase to a phase in which this symmetry is broken. However, while the \PT-symmetry of the Hamiltonian itself does not change along the transition, the behavior of the states does.

Considering operators whose eigenvalues are real or complex conjugate, one can assume that their adjoints are connected to them  by a similarity transformation 
\begin{equation}
	A^{\dagger}=\eta A\eta^{-1},  
	\label{11}
\end{equation}
in which $\eta$ is a Hermitian operator. Operators satisfying this condition have been defined as belonging to the class of pseudo-Hermitian operators \cite{Mostafazadeh2002a,*Mostafazadeh2002b,*Mostafazadeh2002c}. This follows from the fact that using the operator 
$\eta$ as a metric, the internal product can be redefined such that quantum mechanics relations can be extended to the case of \PT-symmetric Hamiltonians \cite{Bender1999,Bender2002,Bender2007}. 

If the matrix $\eta^{1/2}$ such that $\eta^{1/2}\eta^{1/2}=\eta$ and its inverse exist, and are Hermitian, then the matrix
\begin{equation}
	K=\eta^{1/2}A\eta^{-1/2}=\eta^{-1/2}\eta A\eta^{-1}	\eta^{1/2}=K^{\dagger}         
	\label{11a}
\end{equation}
is Hermitian and, therefore, shares with $A$ the same set of eigenvalues. In this case, all of its eigenvalues are real and we have a \PT-symmetric operator in the unbroken phase.

Since the beginning of the studies of \PT-symmetric systems there was an interest in investigating random matrix ensembles to model properties of this kind of Hamiltonians. This comes naturally as symmetries, such as time reversal and rotational, plays an important role in RMT. Several ensembles already have been proposed \cite{Bohigas2013a,Jain2006,Srivastava2012,Marinello2016a} but here we focus on the recently introduced ensemble of pseudo-Hermitian Gaussian matrices \cite{Marinello2016d,Marinello2017} described in the next section. 

It is well known that spectral statistics plays a central role in RMT studies. As a matter of fact, one reason for the success of random matrix models comes from the impact the properties of their spectra had in the characterization of the manifestations of chaos in quantum mechanics \cite{Bohigas1984a,*Bohigas1984b}. This poses the question if specific spectral properties can be associated to the \PT-symmetry or more generally to the pseudo-Hermitian class of operators. Considering the case of unbroken symmetry,  Eq. \eqref{11a} shows that Hermitian and non-Hermitian matrices share the same set of eigenvalues. This suggests that pseudo-Hermiticity or equivalently \PT-symmetry does not seem to induce, in this case, any specific spectral property. On the other hand, when the system is such that it can undergo the phase transition, some universality behavior may be present. 

As the transition proceeds, the spectrum splits in eigenvalues that remain in the real axis while others, in conjugate pairs, evaporate into the complex plane. Repulsion among eigenvalues in the complex plane has already been matter of studies in the case of the eigenvalues of the Ginibre ensemble \cite{Ginibre1965,Mehta2004} and of non-Hermitian normal matrices \cite{Haake2010,Oas1997}. Furthermore, it has been reported an universal cubic repulsion between complex eigenvalues of normal matrices. So, here we are extending those investigations to the case of complex eigenvalues of pseudo-Hermitian operators. 

Turning now to the real ones, it is reasonable to consider that they form a kind of an incomplete sequence of levels with a reduced repulsion among them. The theory of randomly incomplete spectra \cite{Bohigas2006} that emerged from the theory of missing levels \cite{Bohigas2004}, recently, has attracted much attention \cite{Deift2017,Berggren2017,Grabsch2017,Graefe2015}. In \cite{Bohigas2006}, it was indeed conjectured that a situation in which levels move away from the real axis would be a realization of a randomly thinned spectra. Eigenvalues of the pseudo-Hermitian ensemble match that hypothesis.

\section{Overview of the pseudo-Hermitian Gaussian ensemble studied}

The classical Gaussian ensembles of matrices are defined by the distribution \cite{Mehta2004} 
\begin{equation*}
	P(H)=Z^{-1}_N \exp\left[-\frac{\beta}{2}\mbox{tr}(H^{\dagger}H)\right] ,
\end{equation*}  
where $H$ is a matrix with elements that can be written as 
\begin{equation}
	H_{mn}=H_{mn}^{0}+iH_{mn}^{1}+jH_{mn}^{2}+kH_{mn}^{3} 
	\label{7}
\end{equation}
with $i^2=j^2 =k^2=ijk=-1$, and is such that $H$ is symmetric, Hermitian or self dual for real, complex and quaternion matrices, respectively. The number of non-zero elements in Eq. \eqref{7} denoted by $\beta$ can be equal to $1$, $2$ or $4$. Therefore, the elements are Gaussian distributed and can be real, complex or quaternion which define, respectively, the Gaussian orthogonal (GOE), unitary (GUE) and sympletic (GSE) classes. Respectively, the matrices of the classical ensembles are diagonalized by Orthogonal, Unitary and Sympletic matrices.

From the matrices of these ensembles, an ensemble of pseudo-Hermtian Gaussian matrices can be constructed as \cite{Marinello2016d,Marinello2017}
\begin{equation}
	A=PHP+QHQ+r(PHQ-QHP), 
	\label{R1}
\end{equation}
where, with $P_{i}=\left|i\right>\left<i\right|$, we have $P=\sum_{i=1}^{M}P_{i} \quad\mbox{and}\quad Q=\sum_{j=M+1}^{N}P_{j}.$
It is easily verified that matrices of this form satisfy the pseudo-Hermiticity condition, Eq. \eqref{11}, with the metric  defined as $\eta = P - Q .$

Another matrix model has been constructed in Refs. \cite{Marinello2016d,Marinello2017} as
\begin{equation}
	A = \sum_{k=1}^{N}P_{k}HP_{k}+\sum_{j> i}r^{s_{ij}} P_{i}HP_{j}+\sum_{j< i} r^{s_{ij}}\cos[(j-i)\pi]P_{i}HP_{j}, \label{R2}
\end{equation}
where $s_{ij}=1/2-\cos[(j-i)\pi]/2$ and $r$ is a real positive parameter. In this case, the metric can still be written as $\eta=P-Q$, with $P=\sum_{i=1}^{[\frac{N+1}{2}]}P_{2i-1} \mbox{    and    }Q=\sum_{j=1}^{[\frac{N}{2}]}P_{2j} , $
where $[.]$ means integer part. In the model defined by Eq. \eqref{R1}, the matrices are made of separated blocks while in the one defined by Eq. \eqref{R2} the constitutive blocks intertwine. It is convenient to refer to the three classes of pseudo-Hermitian matrices of the ensemble as the pHGOE, the pHGUE and the pHGSE according to the real, complex and quaternion nature of their elements.

\section{Spectral statistics}

\begin{figure}[ht]
	{\includegraphics*[width=0.30\textwidth]{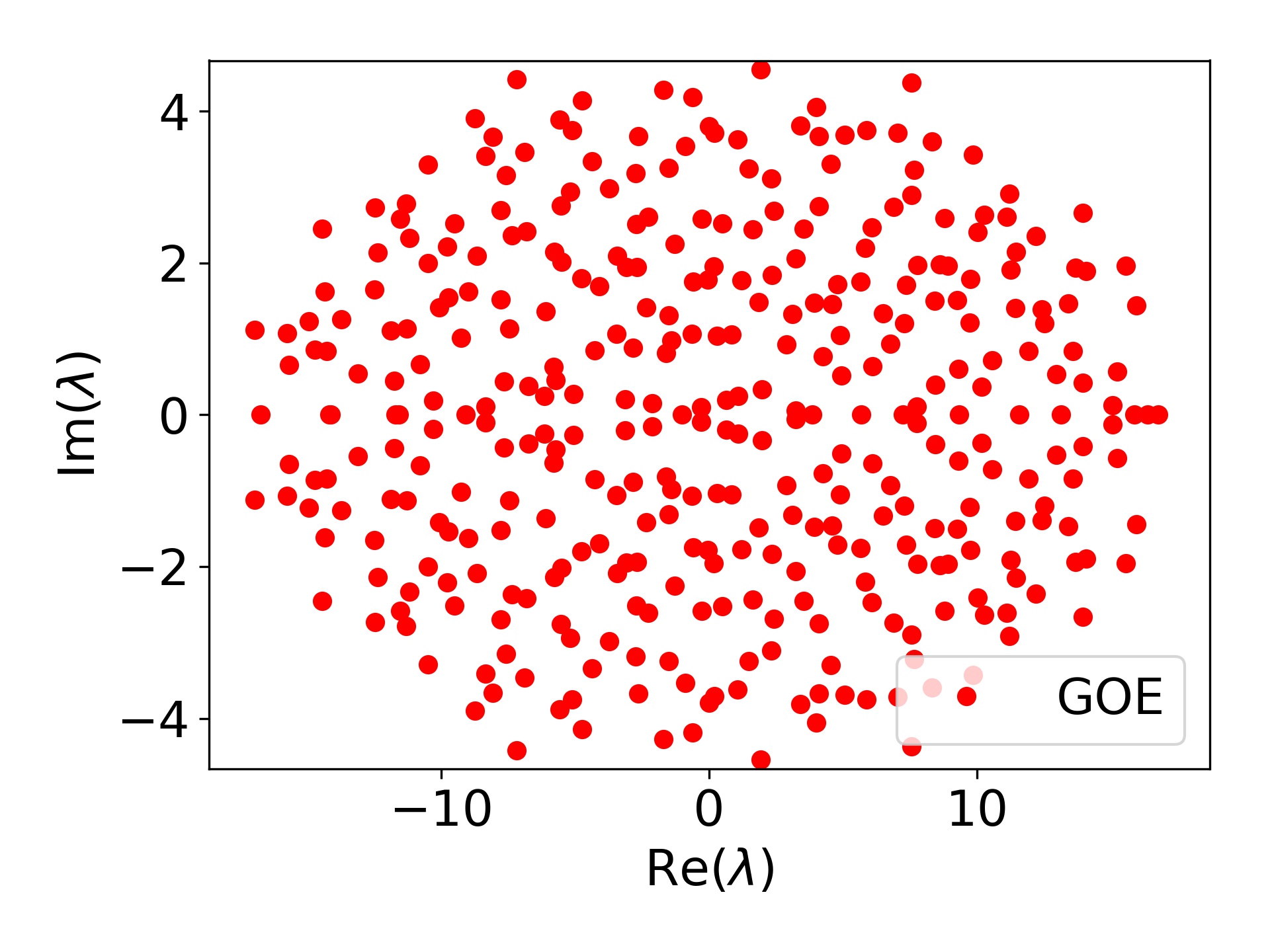}}
	{\includegraphics*[width=0.30\textwidth]{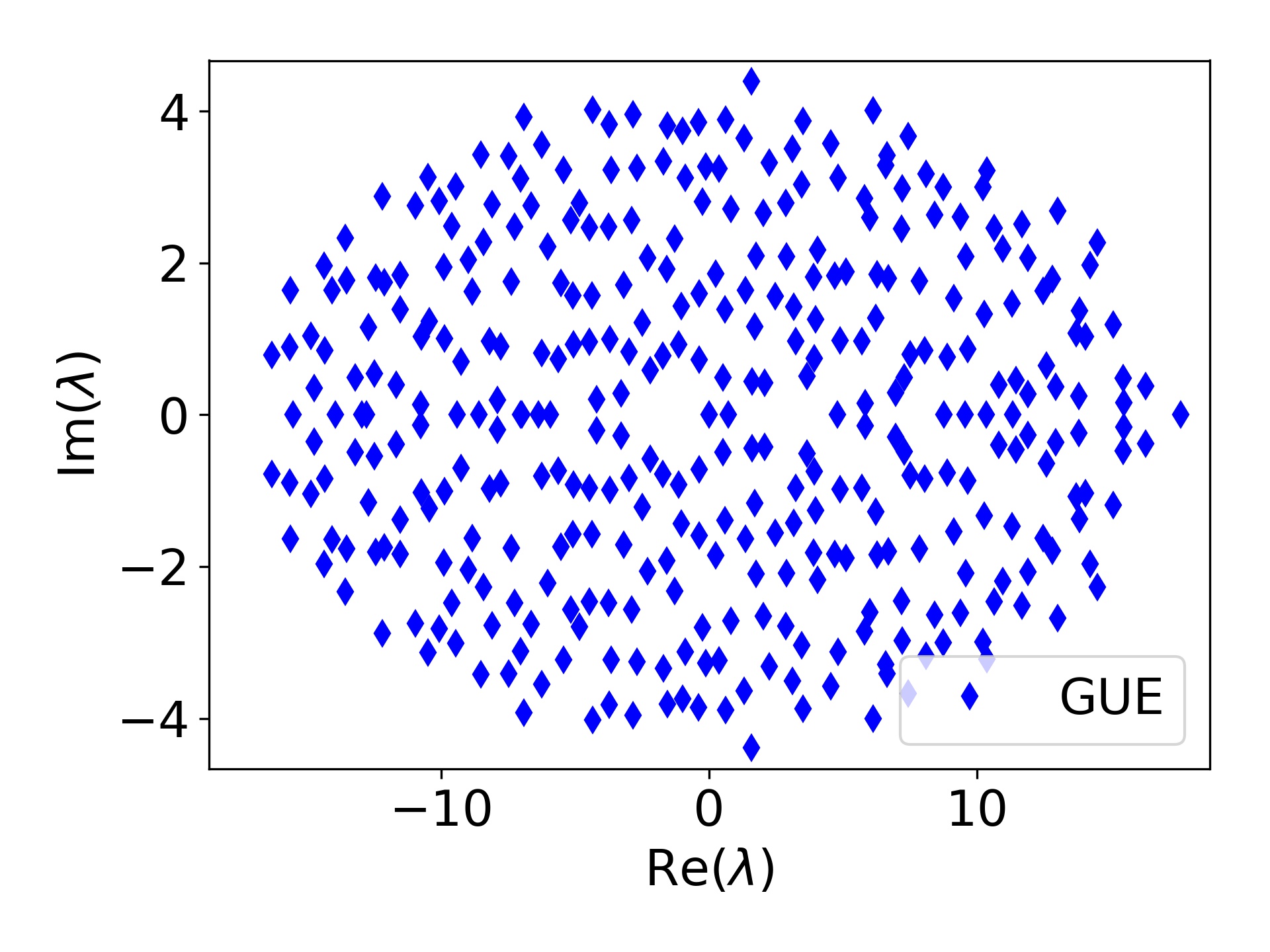}}
	{\includegraphics*[width=0.30\textwidth]{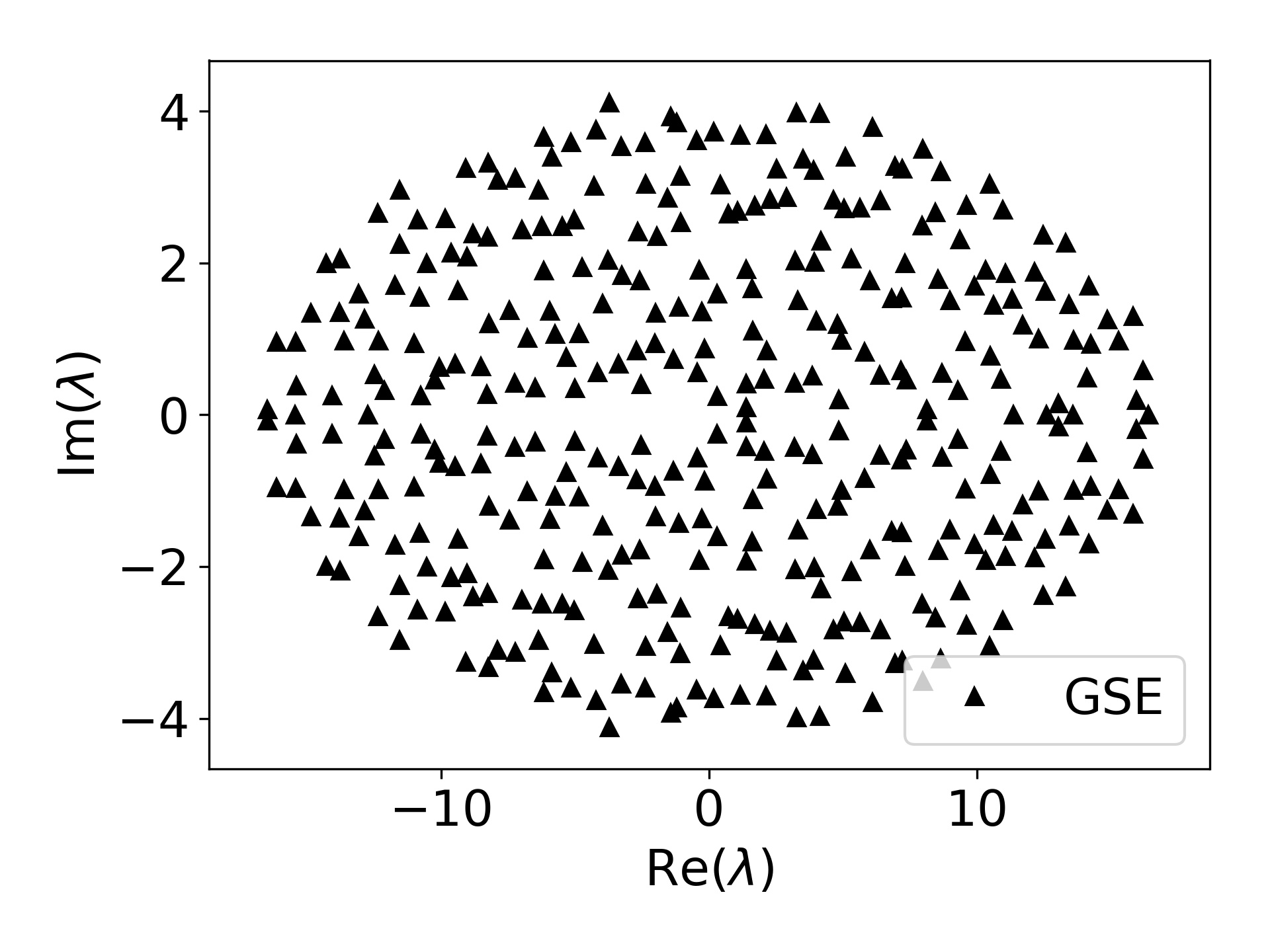}}
	\caption{\label{fig:ellipsis} Real and complex eigenvalues for a single sample matrix of the three classes of pseudo-Hermitian Gaussian matrices for $N = 360$, $M = 180$ and $r = 0.5$.}
\end{figure}

To study the  spectral properties of the pseudo-Hermitian ensembles we use the model defined by Eq. \eqref{R1} and restrict to the case of matrices of even size. Besides the parameter $r$ that fixes the elliptical behavior of the spectra \cite{Marinello2016d} - which will be discussed below - the size $M$ of the smaller block also is a parameter of the model. It more or less fixes the number of pairs of eigenvalues that leave, on average, the real axis. We remark that, in particular, for $M=N/2$ it gives the same results using Eq. \eqref{R2}, and that this is the case for which, as $r$ increases, eigenvalues eventually all evaporate into the complex plane \cite{Marinello2016d}. The effect of thinning on the real spectrum can be assessed by varying the size $M$ or the parameter $r$. In Fig. \ref{fig:ellipsis}, the real and complex conjugate eigenvalues are displayed for a sample matrix for each of the three classes of matrices for $M$ one third of their size. It is clear that the way the cloud of the complex eigenvalues fills the ellipsis is the result of the interplay between confinement and repulsion among them and may in principle be distinct for each case of $\beta$, even though our choice of variance for the elements implies that they fall on the same ellipsis. In the following sections, we shall present numerical case studies that confirm the distinct behavior of each of those cases. 

\begin{figure}[ht]
	{\includegraphics*[width=0.30\textwidth]{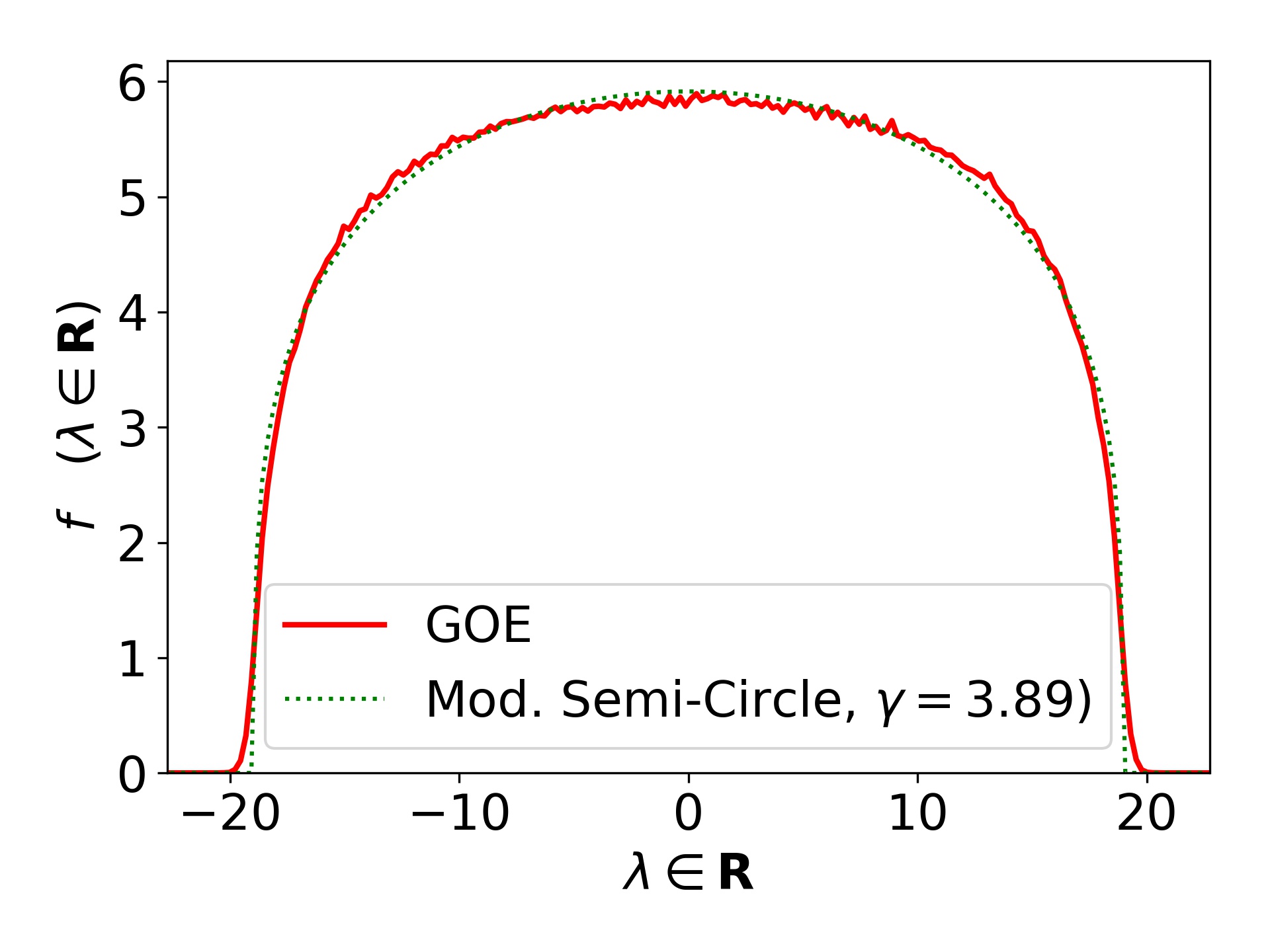}}
	{\includegraphics*[width=0.30\textwidth]{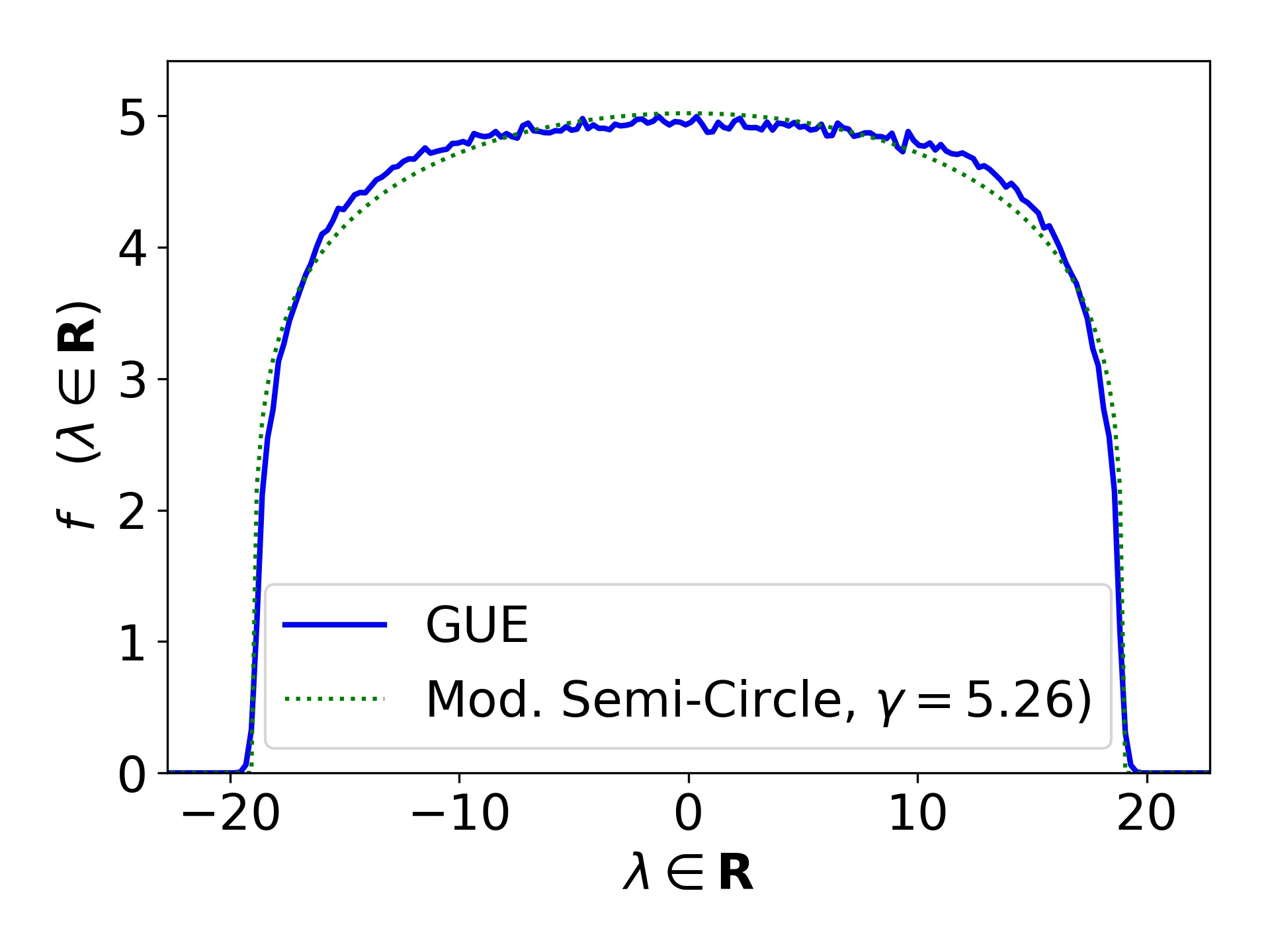}}
	{\includegraphics*[width=0.30\textwidth]{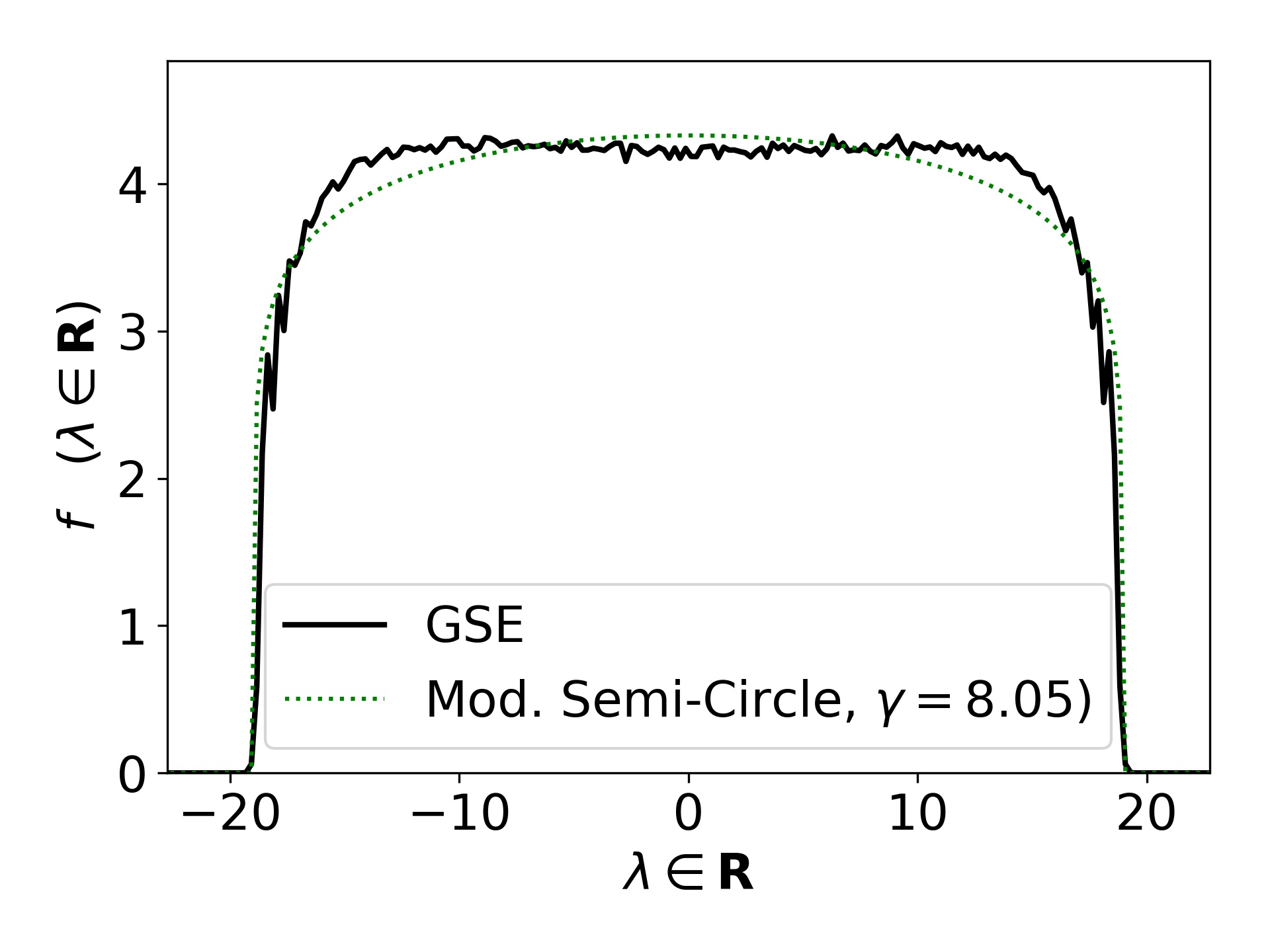}}
	\caption{\label{fig:rho:lowr} Density of real eigenvalues for the indicated values of the parameters	fitted using the modified semi-circle for the $\gamma$ parameter as indicated in the figure, calculated from a sample of $2\times 10^4$ matrices from the pseudo-Hermitian ensembles, with $N = 360$, $M = 180$ and $r = 0.05$. Mean number of real eigenvalues observed was $n_{\mbox{\tiny phGOE}}=195.3$, $n_{\mbox{\tiny phGUE}}=171.2$ and $n_{\mbox{\tiny phGSE}}=152.8$.}
\end{figure}

\begin{figure}[ht]
	{\includegraphics*[width=0.30\textwidth]{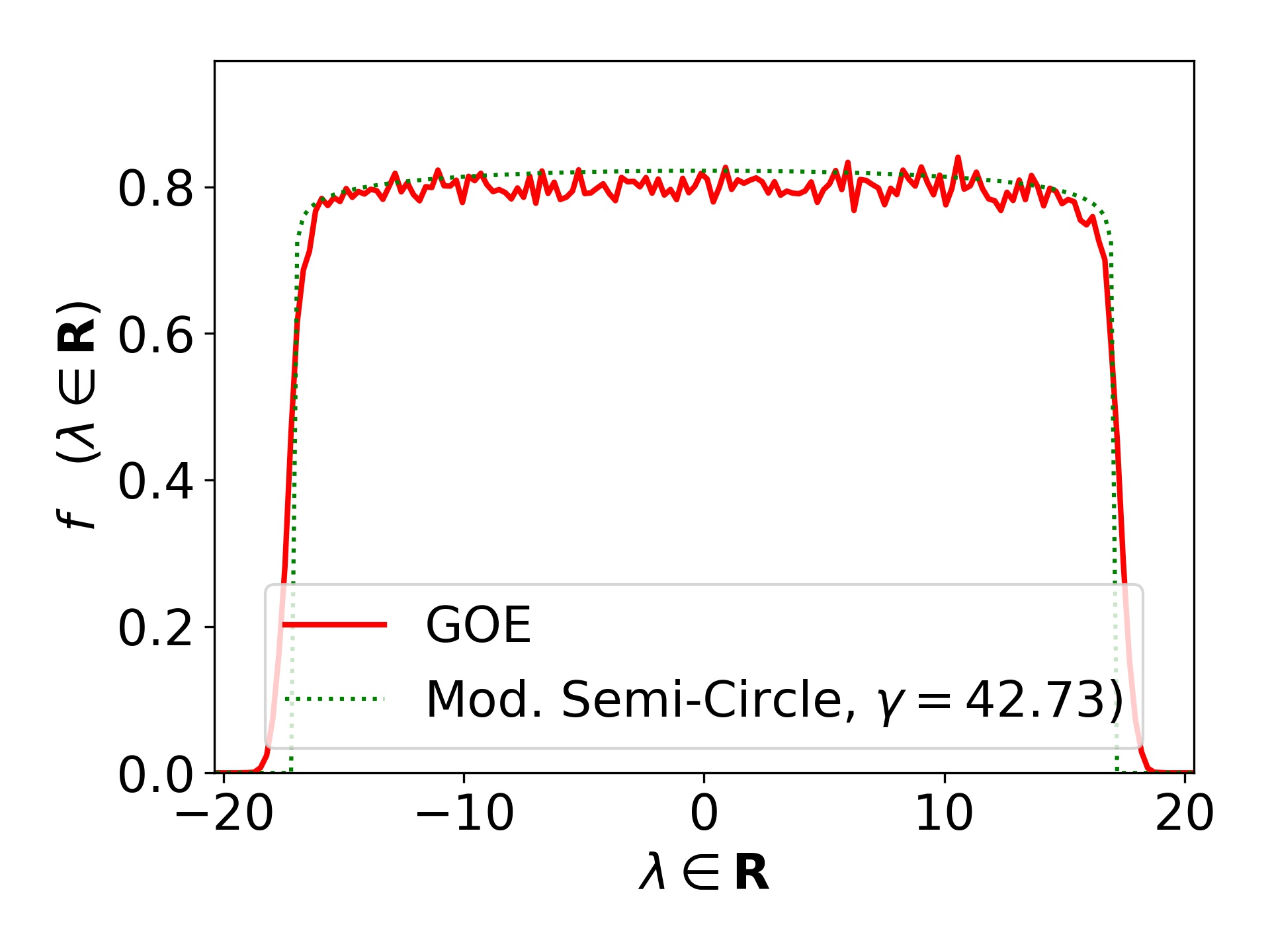}}
	{\includegraphics*[width=0.30\textwidth]{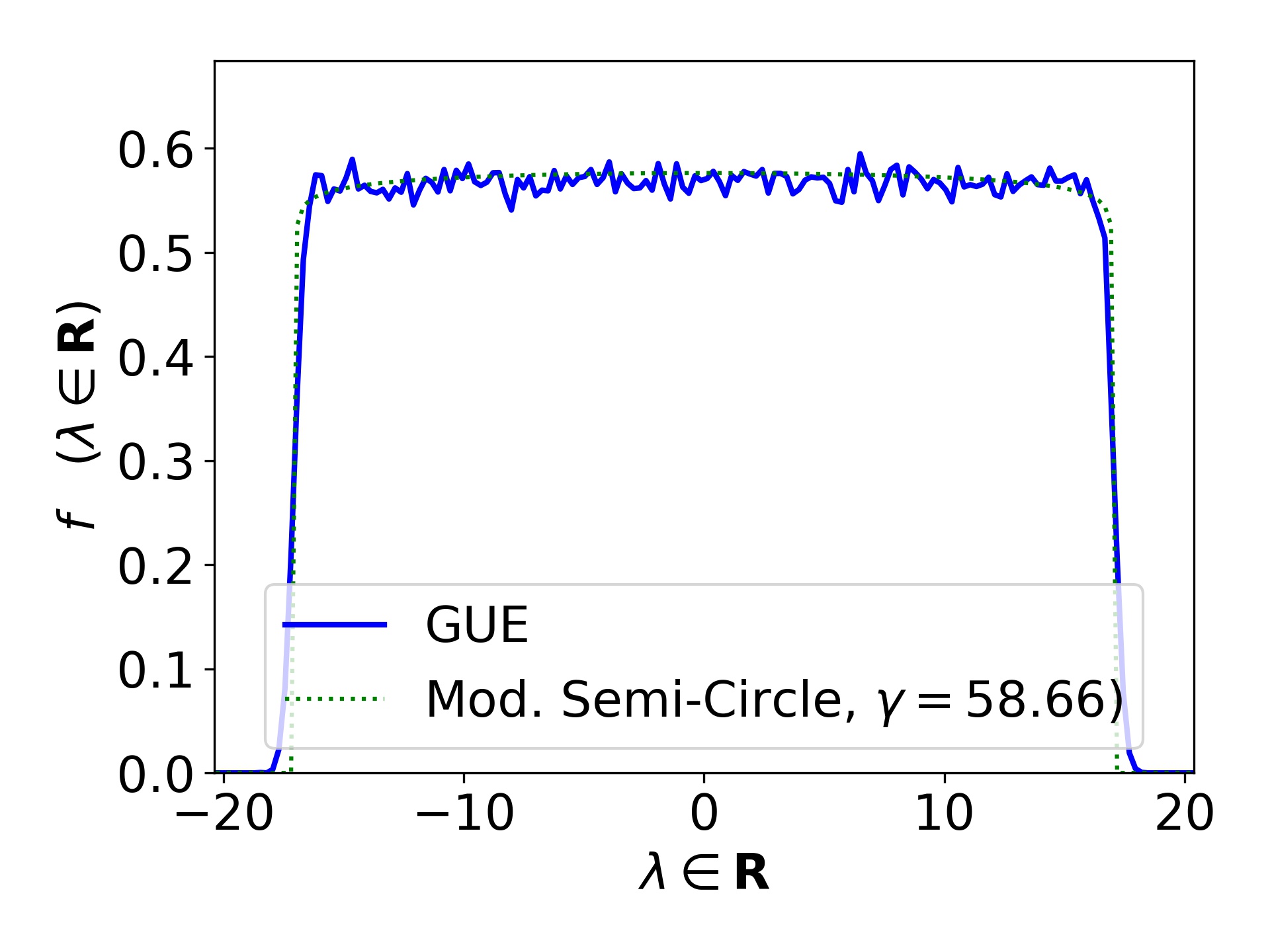}}
	{\includegraphics*[width=0.30\textwidth]{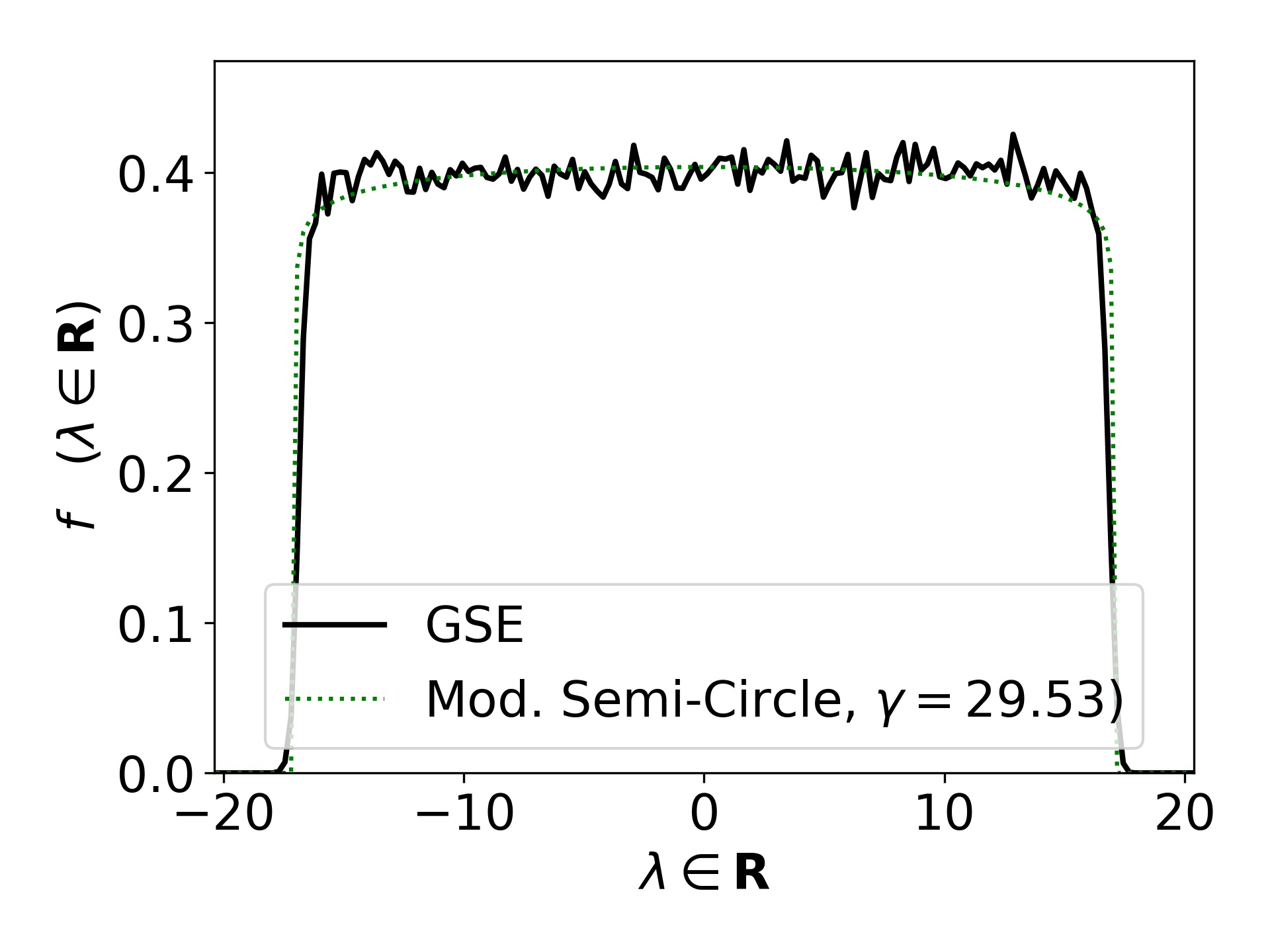}}
	\caption{\label{fig:rho:medr} Density of real eigenvalues for the indicated values of the parameters	fitted using the modified for the $\gamma$ parameter as indicated in the figure, calculated from a sample of $2\times 10^4$ matrices from the pseudo-Hermitian ensembles, with $N = 360$, $M = 180$ and $r = 0.50$. Mean number of real eigenvalues observed was $n_{\mbox{\tiny phGOE}}=27.5$, $n_{\mbox{\tiny phGUE}}=19.4$ and $n_{\mbox{\tiny phGSE}}=13.4$.}
\end{figure}

\begin{figure}[ht]
	{\includegraphics*[width=0.30\textwidth]{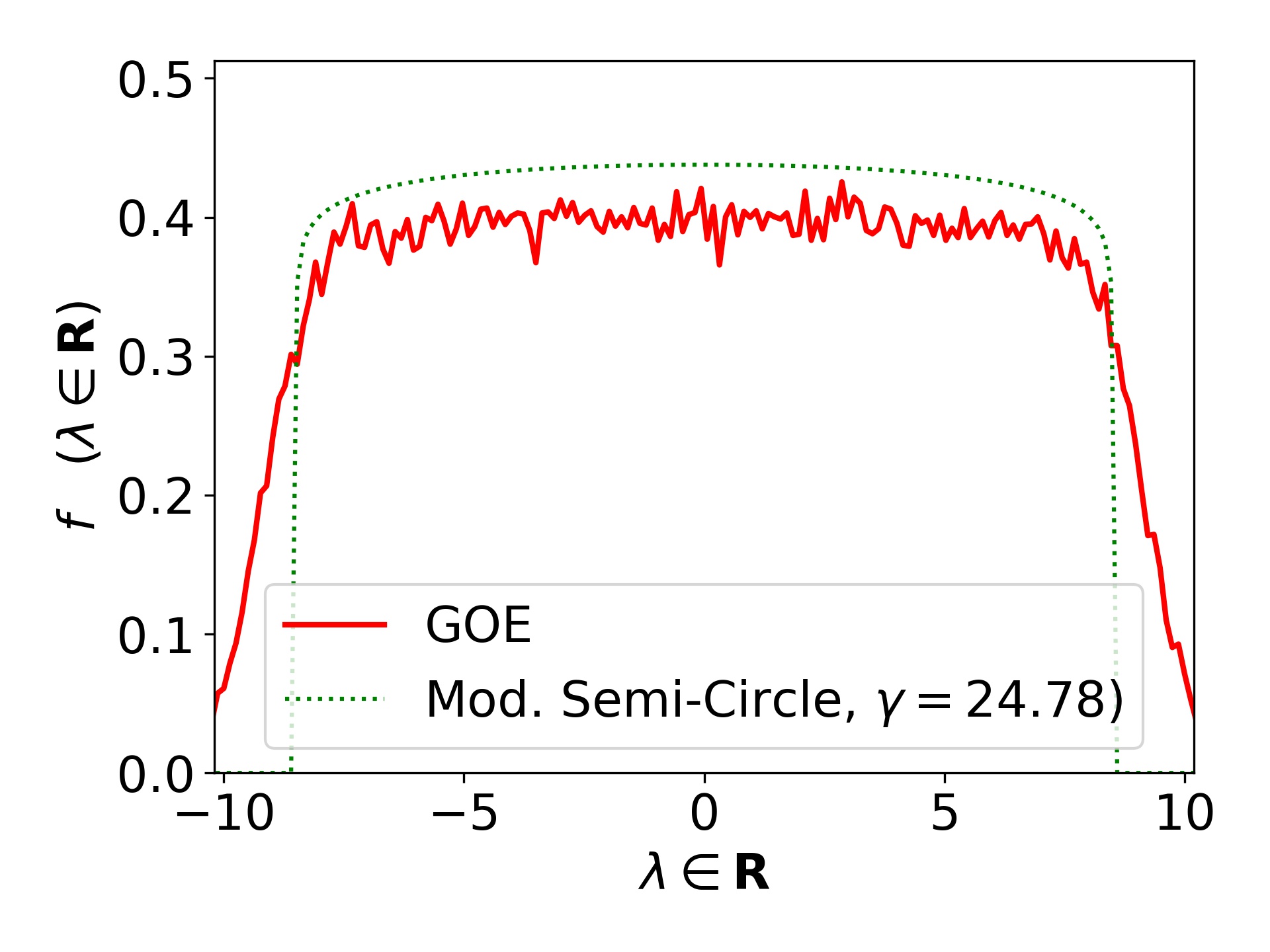}}
	{\includegraphics*[width=0.30\textwidth]{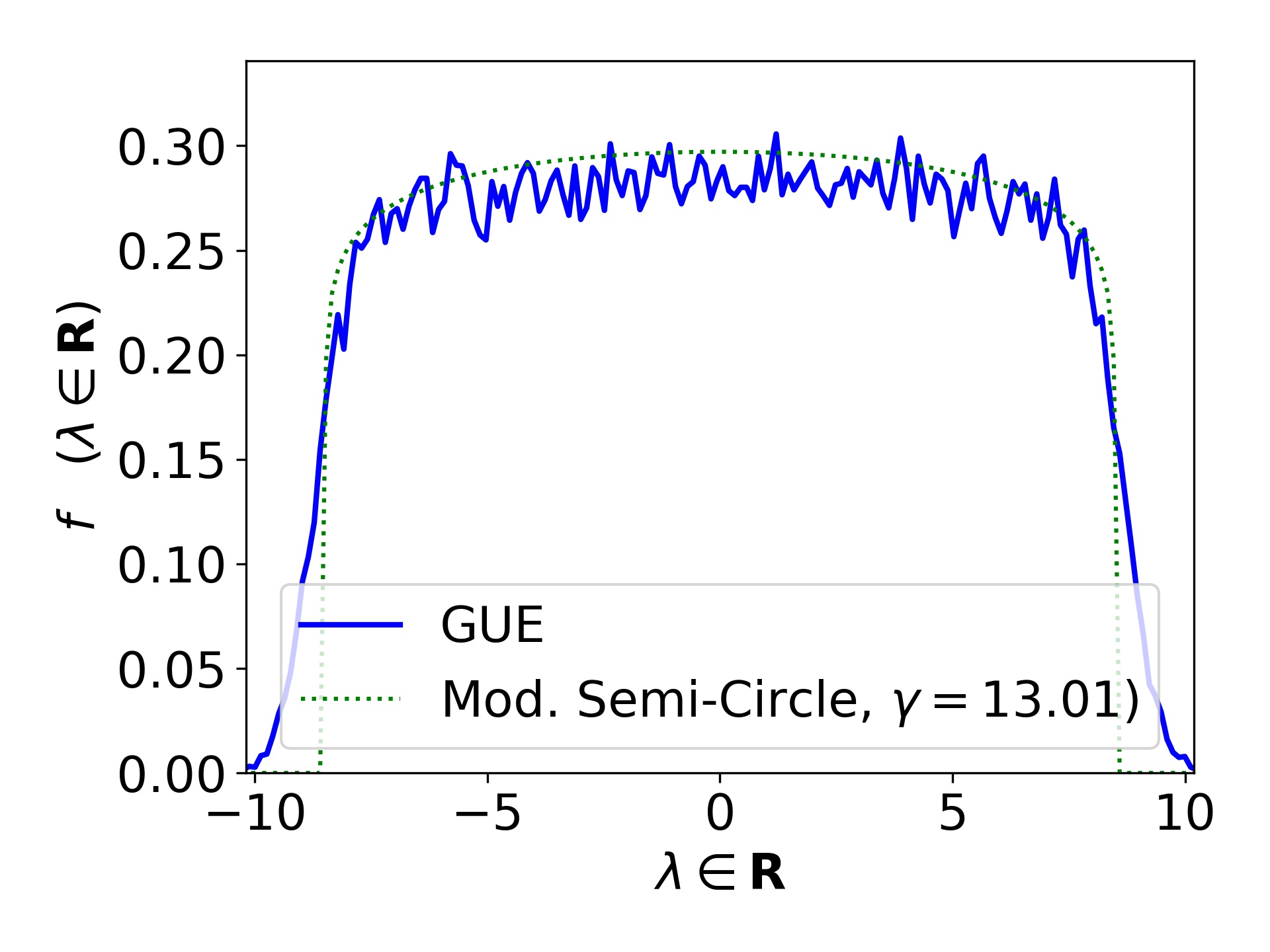}}
	{\includegraphics*[width=0.30\textwidth]{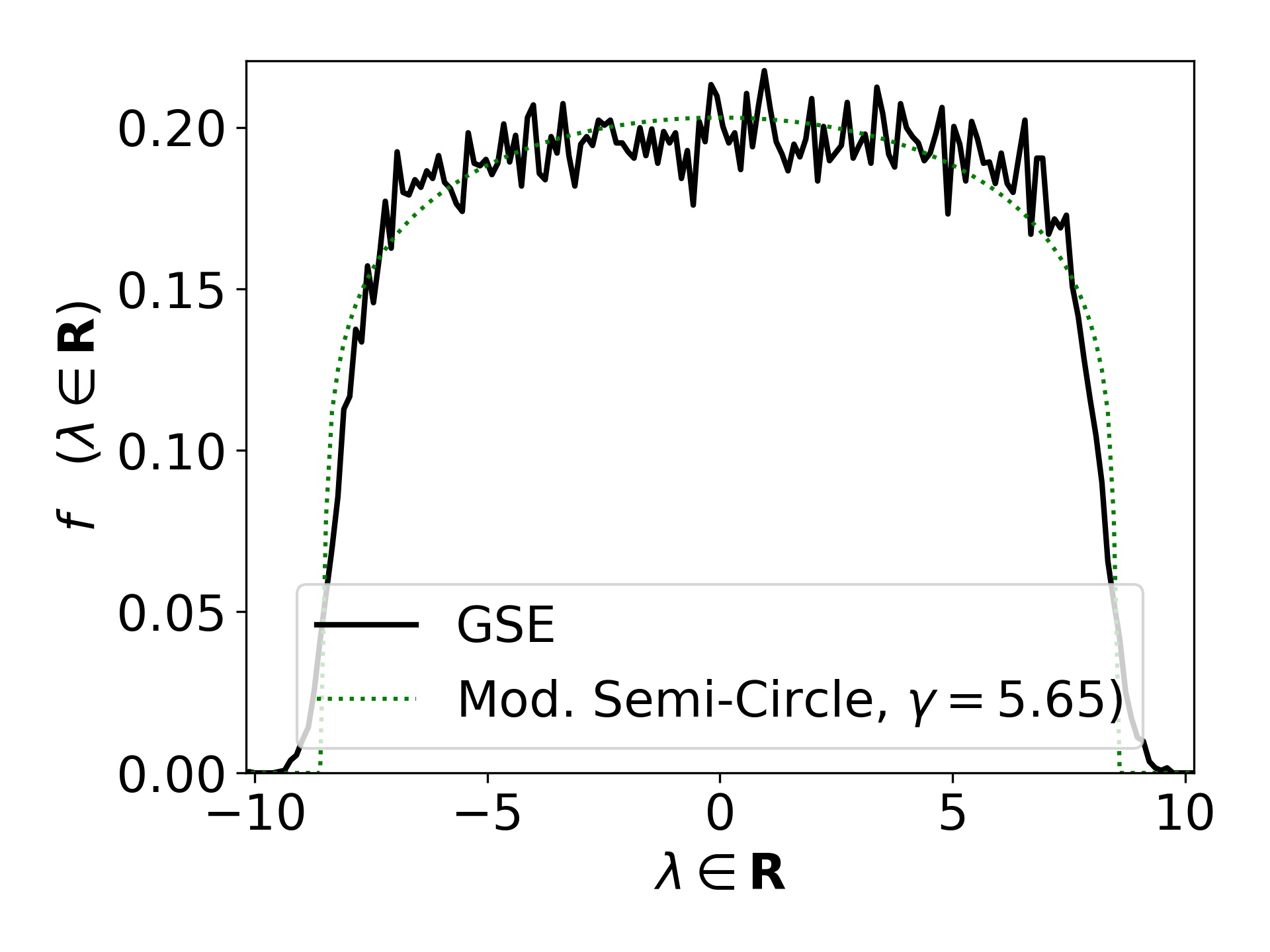}}
	\caption{\label{fig:rho:highr} Density of real eigenvalues for the indicated values of the parameters	fitted using the modified semi-circle for the $\gamma$ parameter as indicated in the figure, calculated from a sample of $2\times 10^4$ matrices from the pseudo-Hermitian ensembles, with $N = 360$, $M = 180$ and $r = 2.00$. Mean number of real eigenvalues observed was $n_{\mbox{\tiny phGOE}}=7.2$, $n_{\mbox{\tiny phGUE}}=4.8$ and $n_{\mbox{\tiny phGSE}}=3.1$.}
\end{figure}

Furthermore, in Figs. \ref{fig:rho:lowr}-\ref{fig:rho:highr} the density of real eigenvalues of the pHGOE, pHGUE and phGSE case is presented, normalized to the average number of real eigenvalues. The dashed green line in those graphs is the fit for the modified semi-circle \eqref{127}, discussed in the following section, which is performed withing the bounds for which that Eq. is real and well-defined, $x \in [-a,+a]$. For values of $r$ close to zero, such as $r = 0.05$ in Fig. \ref{fig:rho:lowr}, the density is beginning to transition from the semi-circle into the new density which is observed for higher values of $r$. In Fig. \ref{fig:rho:medr} we may observe that the density rapidly plateaus even for $r = 0.5$. This is still observed in Fig. \ref{fig:rho:highr}. It is notable, however, that increasing the parameter $r$ causes the number of remaining real eigenvalues to drop sharply, and some border effects begin to appear.

\subsection{Real eigenvalues statistics}

We start by investigating the density of the fraction of eigenvalues that remains on the real axis. We have found that for $M=N/2$ this density approximately can be fitted with the modified semi-circle 
\begin{equation}
	\rho(x)=\frac{\Gamma\left(\frac{3}{2}+\frac{1}{\gamma}\right)}{a^{1+\frac{2}{\gamma}}\sqrt{\pi}\Gamma\left(1+\frac{1}{\gamma}\right)}(a^2-x^2)^{1/\gamma}
	\label{127},
\end{equation}
where $\frac{1}{2} \leq \gamma <\infty$ and $a = \sqrt{\frac{N}{1+r^2}}$ and $C$ is a normalization constant. This means that, as $r$ increases the density goes from the semi-circle to an uniform distribution. From this density, the unfolded spectrum can be derived using the cumulative function 
\begin{equation*}
	N(x)=C\int_{-R}^{x}dt(a^2-t^2)^{1/\gamma} = \frac{1}{2} + \frac{\Gamma\left(\frac{3}{2}+\frac{1}{\gamma}\right)}{a\ \sqrt{\pi}\  \Gamma\left(1+\frac{1}{\gamma}\right)} {}_2F_1\left(\frac{1}{2},-\frac{1}{\gamma};\frac{3}{2};\frac{x^2}{a^2}\right).
\end{equation*}
where ${}_2F_1(a_1,a_2;b_1;z)$ is the hypergeometric function. 

However, as a plateau in the density shows up quickly, that is, for relatively small values of $r$, by discarding eigenvalues close to the edges, the density can be better treated as uniform. As a consequence, the average spacing is constant and easily can be made equal to one. It is notable that for the central region of the spectra there is a fair agreement with the fit for the modified semi-circle, the general exception being the phGSE in both $r = 0.05$ and $r = 2.00$ cases, and the constant density approximation is a good descriptor - far from the spectral edges - for all but the phGOE and phGUE cases of $r = 0.05$. The difficulty in obtaining good statistics for the real eigenvalues of the $r = 2.00$ is also evident, although the constant approximation seems to be a better descriptor for the density far from the spectral edges. 

\begin{figure}[ht]
	\includegraphics*[width=0.30\linewidth]{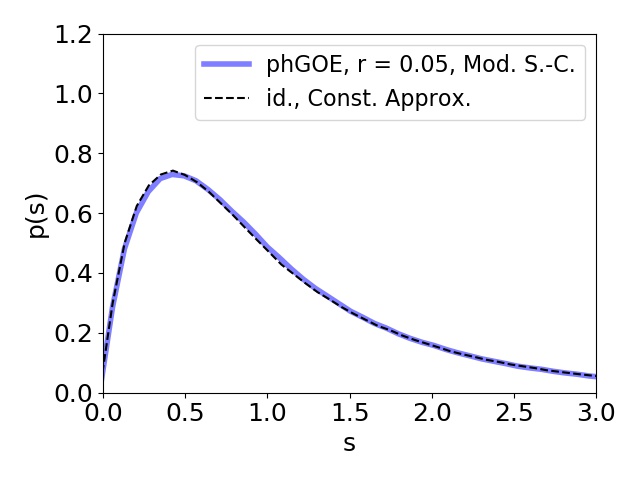}
	\includegraphics*[width=0.30\linewidth]{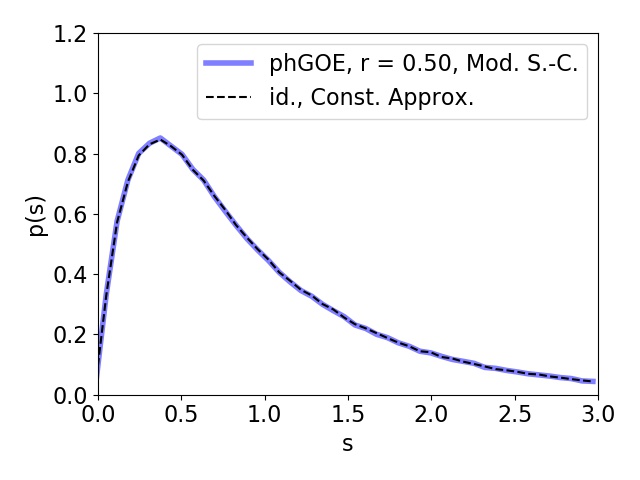}
	\includegraphics*[width=0.30\linewidth]{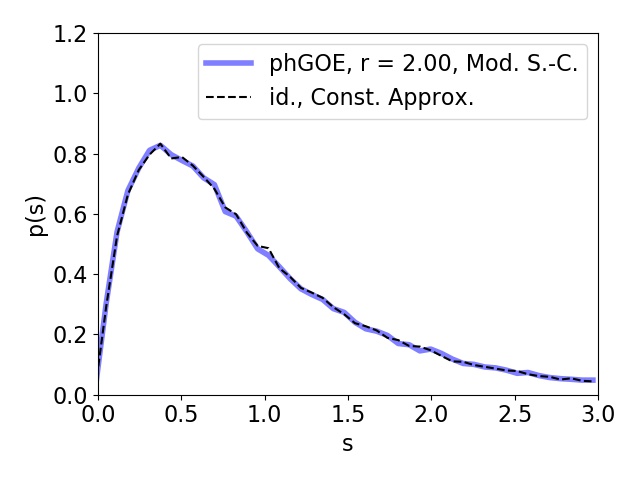}
	\includegraphics*[width=0.30\linewidth]{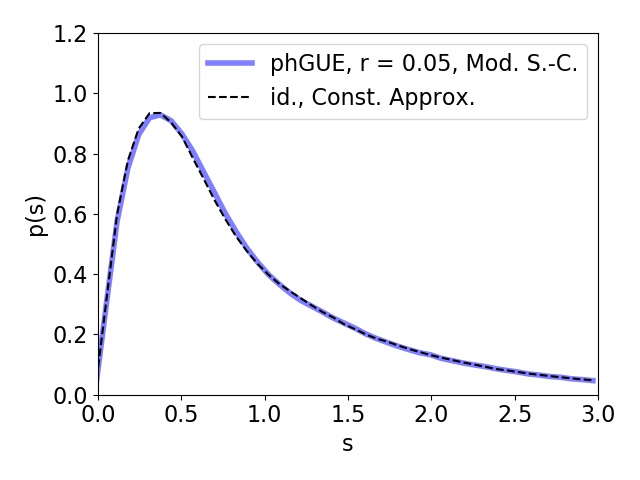}
	\includegraphics*[width=0.30\linewidth]{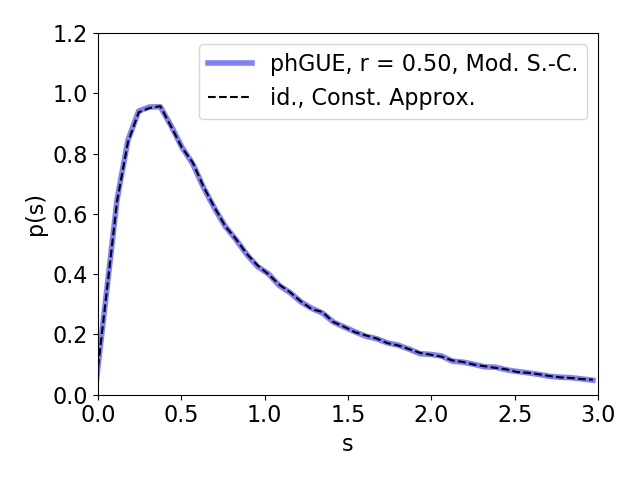}
	\includegraphics*[width=0.30\linewidth]{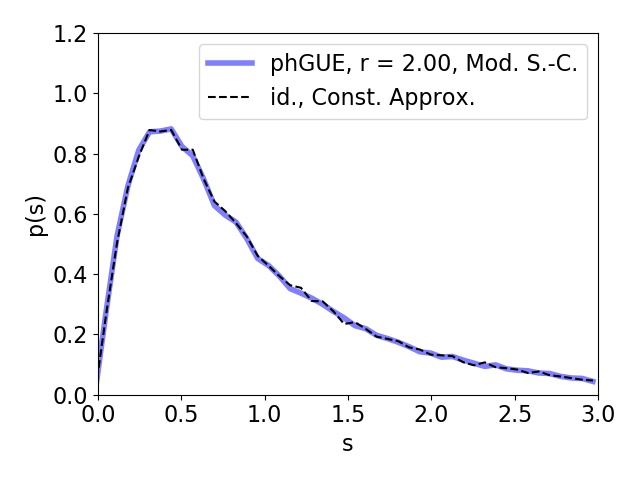}
	\includegraphics*[width=0.30\linewidth]{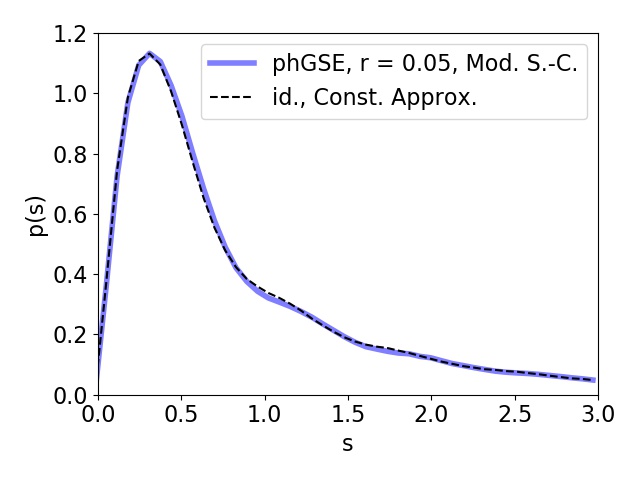}
	\includegraphics*[width=0.30\linewidth]{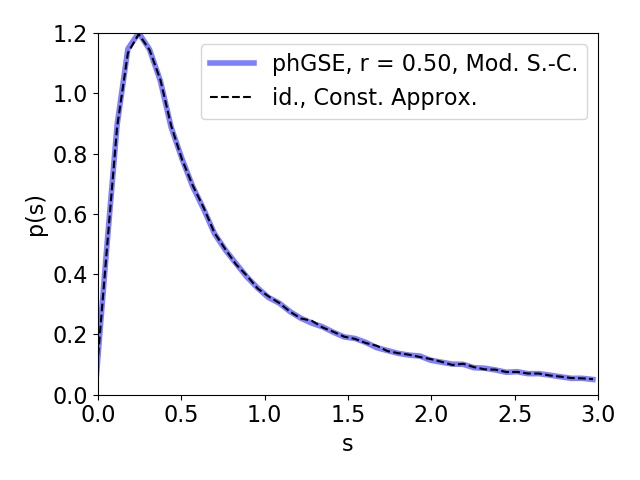}
	\includegraphics*[width=0.30\linewidth]{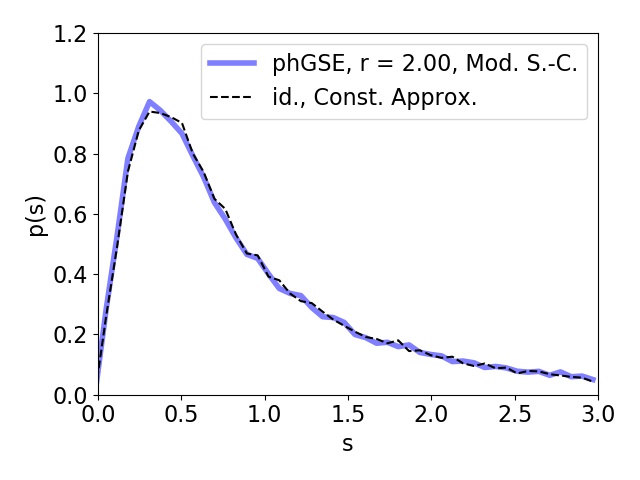}
	\caption{\label{fig:spccompare} Real eigenvalue spacing for the three pseudo-Hermitian cases, for the sample of $2\times 10^4$ matrices of size $N = 360$, with $M = N/2$. The parameter $r$ and corresponding ensemble are as indicated in each plot, with the blue line denoting the modified semi-circle fit, and the dashed black line denoting the constant matrix approximation.}
\end{figure}

Therefore, calculating the spacing has two routes, namely using the modified semi-circle, or using the constant approximation. In Fig. \ref{fig:spccompare} we present the comparison between the two approximations for the three aforementioned cases of $r = 0.05$, $r = 0.5$ and $r = 2.0$. There are notable differences in the phGOE and phGUE for the $r = 0.05$ cases, for which the constant approximation shows a consistent underestimation of the spacing. For the remaining cases, however, there is considerable agreement in the calculated spacings, although good statistics for the $r = 2.00$ case proved again difficult to obtain.

\begin{figure}[ht]
	{\includegraphics*[width=0.50\linewidth]{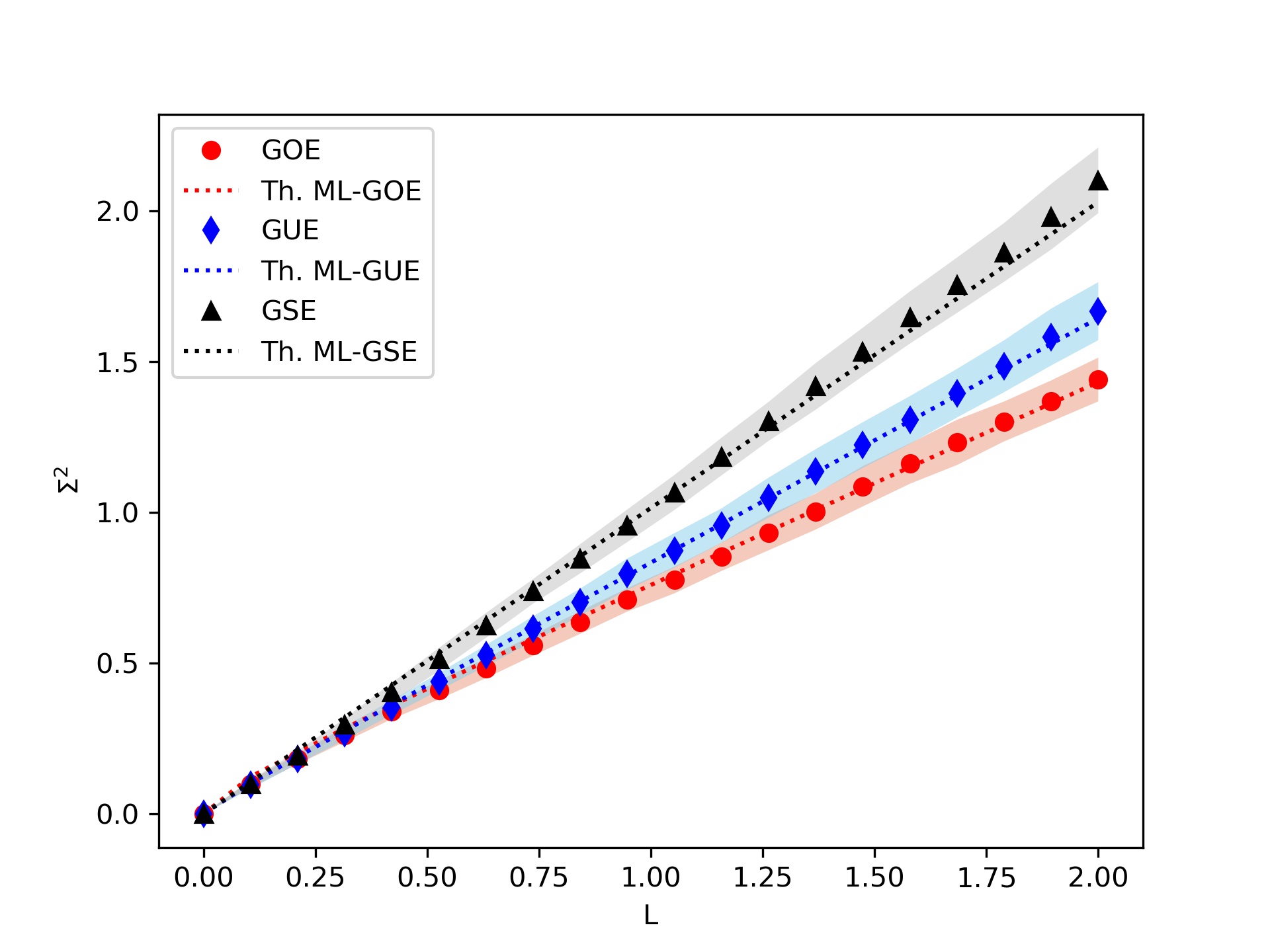}}
	\caption{\label{fig:NV} Number variance calculated for the pseudo-Hermitian ensembles, for a sample of $2 \times 10^4$ matrices, with $N = 360$, $M = 180$ and $r = 0.5$, divided in 40 subgroups of 500 matrices. Markers represent the corresponding average behavior of the sample data, whereas the colored band depicts one standard deviation above and below the average. The dotted line shows the fitted curve corresponding Eq. \eqref{eq:NV}. The equivalent missing level remaining fraction $f$ was found to be $f_{\mbox{\tiny phGOE}} = 0.344(12)$, $f_{\mbox{\tiny phGUE}} = 0.191(04)$ and $f_{\mbox{\tiny phGSE}} = -0.014(07)$, within the statistical uncertainty of the Levenberg–Marquardt least square algorithm used, and the unfolding was calculated using the constant approximation. }
\end{figure}

This allows us to calculate in a straightforward manner, from the eigenvalue sample, the number variance $\Sigma^2 (L)$, defined as the variance of the number of eigenvalues in the interval from $-L/2$ to $L/2$. For random matrices of the classical Gaussian ensembles from which only a fraction $f$ of eigenvalues is left remaining, the expected behavior, from Ref. \cite{Bohigas2004}, is:
\begin{equation}
\Sigma^2(L) = (1-f) L + f^2\Sigma^2_G\left(\frac{L}{f}\right)
\label{eq:NV}
\end{equation}
where $\Sigma^2_G$ is the number variance for the corresponding Gaussian ensemble. This allows us to compare how the number variance of the pseudo-Hermitian ensembles studies here compare to the missing level theory. This is depicted in Fig. \ref{fig:NV}, for which the number variance is presented and the parameters were fitted. Using the sample of $2 \times 10^4$ matrices with $N = 360$, $M =  180$ and $r = 0.5$, divided in 40 sub-samples of 500 matrices. The number variance was obtained, and presented with a band depicting up to one standard deviation above and below the average observed value. 
While the phGOE and phGUE were well represented by the missing levels model, the real eigenvalues in phGSE become so thinned that they approach the behavior expected of the Poisson case \cite{Bohigas2004,Mehta2004}, rather than being compatible to the any possible value of missing levels. Nonetheless, the fit can still be calculated, and has therefore been included for comparison to the other two cases.

\begin{figure}[ht]
	{\includegraphics*[width=0.50\linewidth]{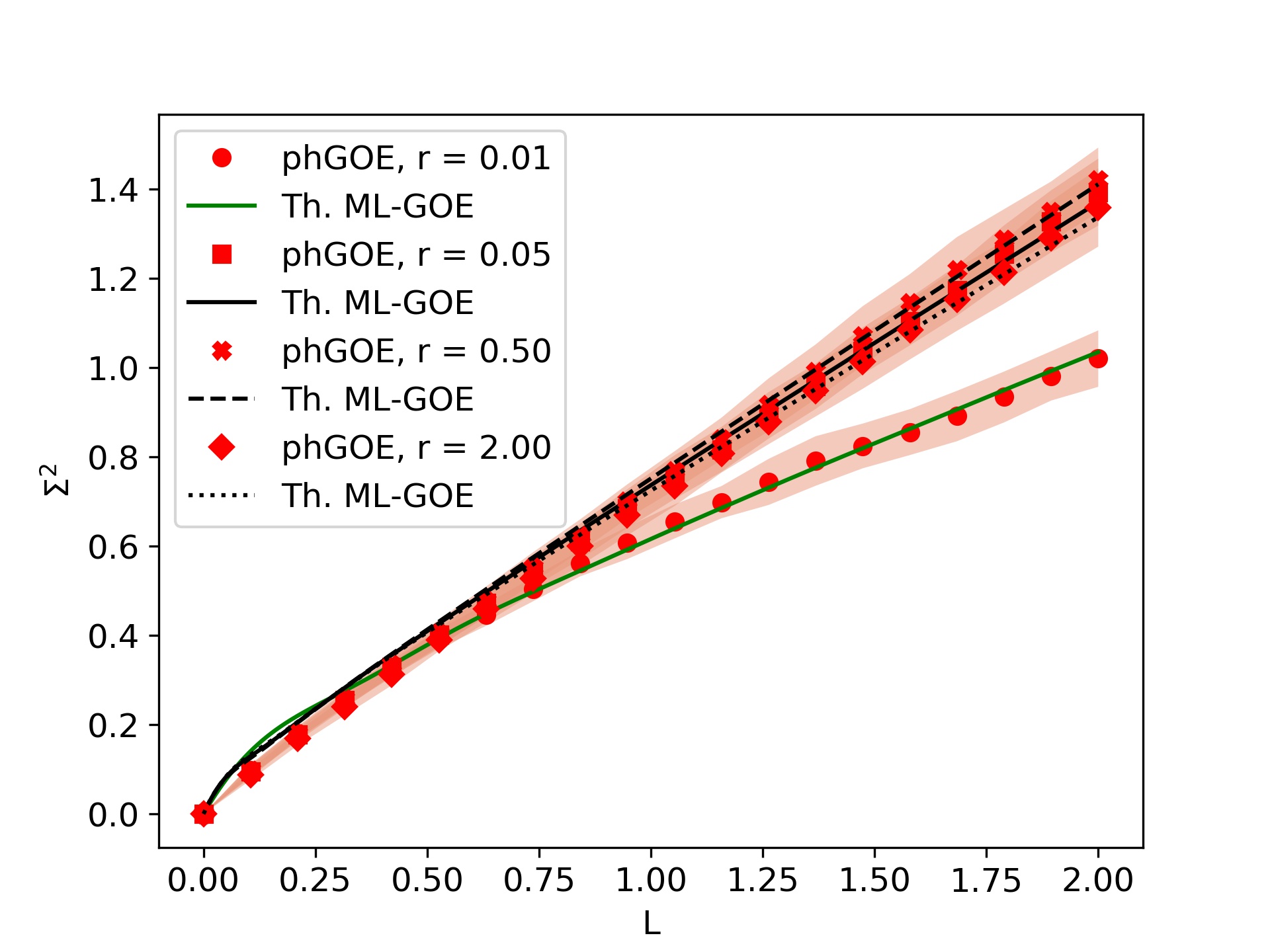}}
	\caption{\label{fig:NV:phGOE} Number variance calculated for the phGOE, for a sample of $2 \times 10^4$ matrices, with $N = 360$, $M = 180$ and $r = 0.01$, $0.05$, $0.5$ and $2.0$, each divided in 40 subgroups of 500 matrices. Markers represent the corresponding average behavior of the sample data, whereas the colored band depicts one standard deviation above and below the average. The solid, dashed and dotted black lines represent parameters $r = 0.05$, $0.5$ and $2.0$. The value obtained for the fraction $f$ is, respectively, $f = 0.638(18)$, $0.387(17)$, $ 0.358(14)$ and $ 0.410(24)$, within the statistical uncertainty of the Levenberg–Marquardt least square algorithm used nd the unfolding was calculated using the modified semi-circle. }
\end{figure}

It is also worthy of note that the three parameters studied present very similar behavior for their number variances. In Fig. \ref{fig:NV:phGOE} we present the comparison between the results of the three parameters $r$ considered above. It is notable that they were all calculated within the margins of statistical error of one another. In the limit of $r \to 0$, we are expected to recover similar behavior to that of the classical GOE number variance \cite{Mehta2004}, and we have therefore included the lower value of $r =0.01$, for which the behavior bears a close qualitative resemblance to that case \cite{Note}. While this seems to suggest the existance of some form of critical point for  which the long-range properties of the real eigenvalues begin to saturate, a complete analysis of this question is beyond the scope of the present paper.

\begin{figure}[ht]
	{\includegraphics*[width=0.50\textwidth]{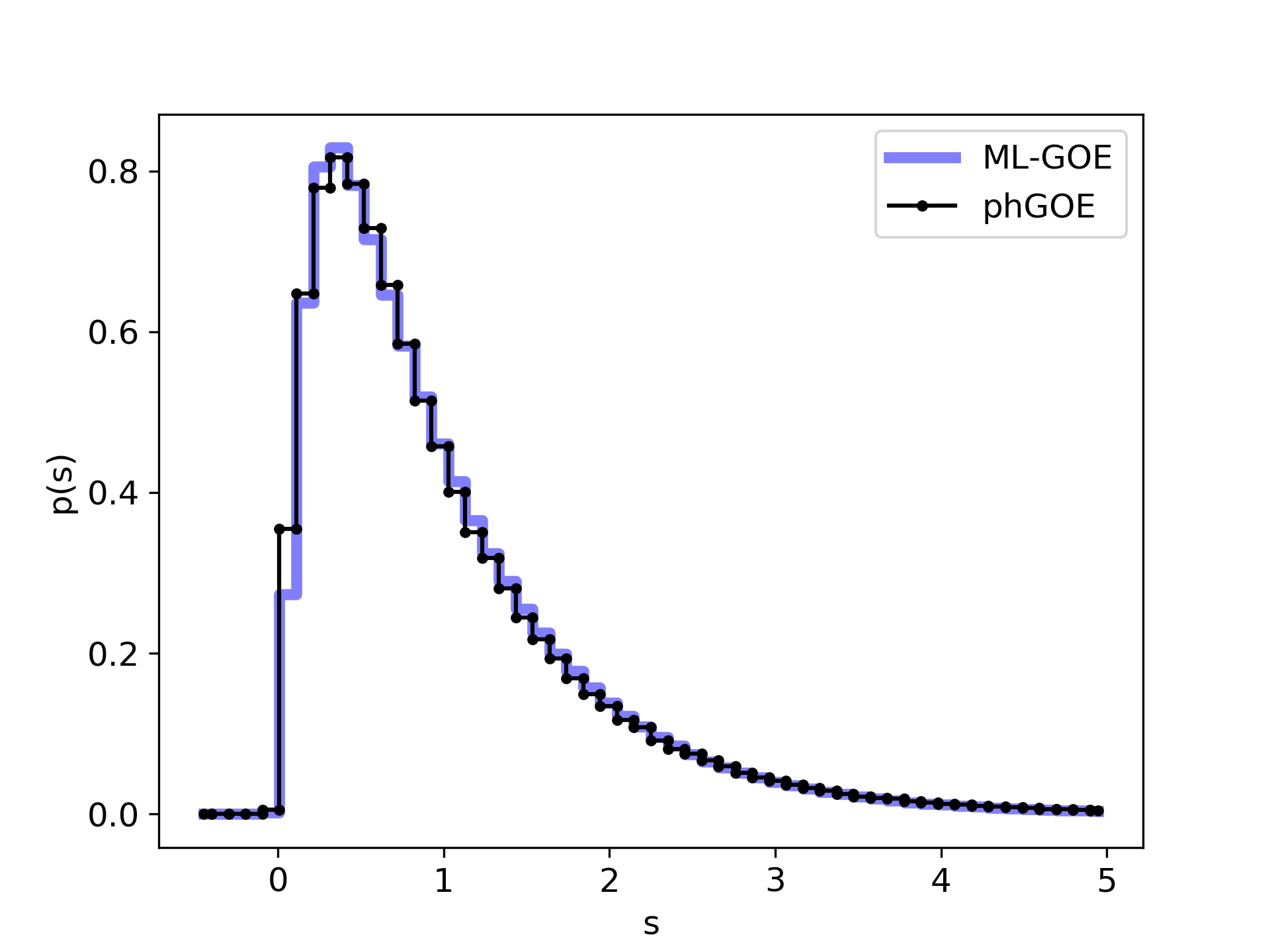}}
	\caption{\label{fig:sp:pHGOE} Black: spacing distribution of the real eigenvalues of the pHGOE ($\beta=1$) with $r=0.50$, with $N = 360$ and $M =90$; blue: spacing distribution of Hermitian GOE spectrum with a random fraction $f=0.31$ of levels remaining. In both cases, a sample size of $2 \cdot 10^4$ was used.}
\end{figure}

\begin{figure}[ht]
	{\includegraphics*[width=0.50\textwidth]{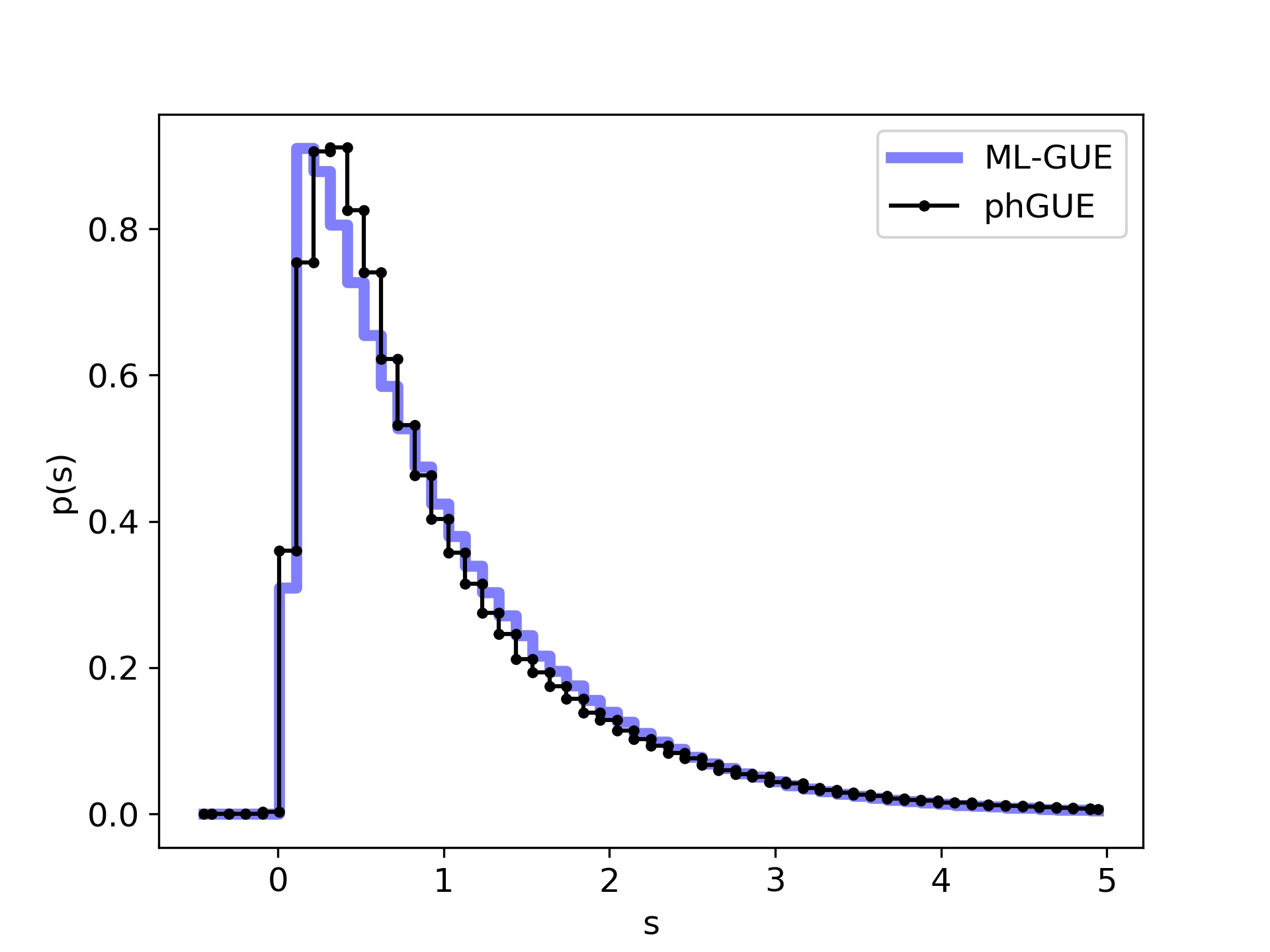}}
	\caption{\label{fig:sp:pHGUE} Black: spacing distribution of the real eigenvalues of the pHGUE ($\beta=2$) with $r=0.50$, with $N = 360$ and $M =90$; blue: spacing distribution of Hermitian GUE spectrum with a random fraction $f=0.18$ of levels remaining. In both cases, the same sample of size of $2 \cdot 10^4$ from Fig. \ref{fig:rho:medr} was used.}
\end{figure}

\begin{figure}[ht]
	{\includegraphics*[width=0.50\textwidth]{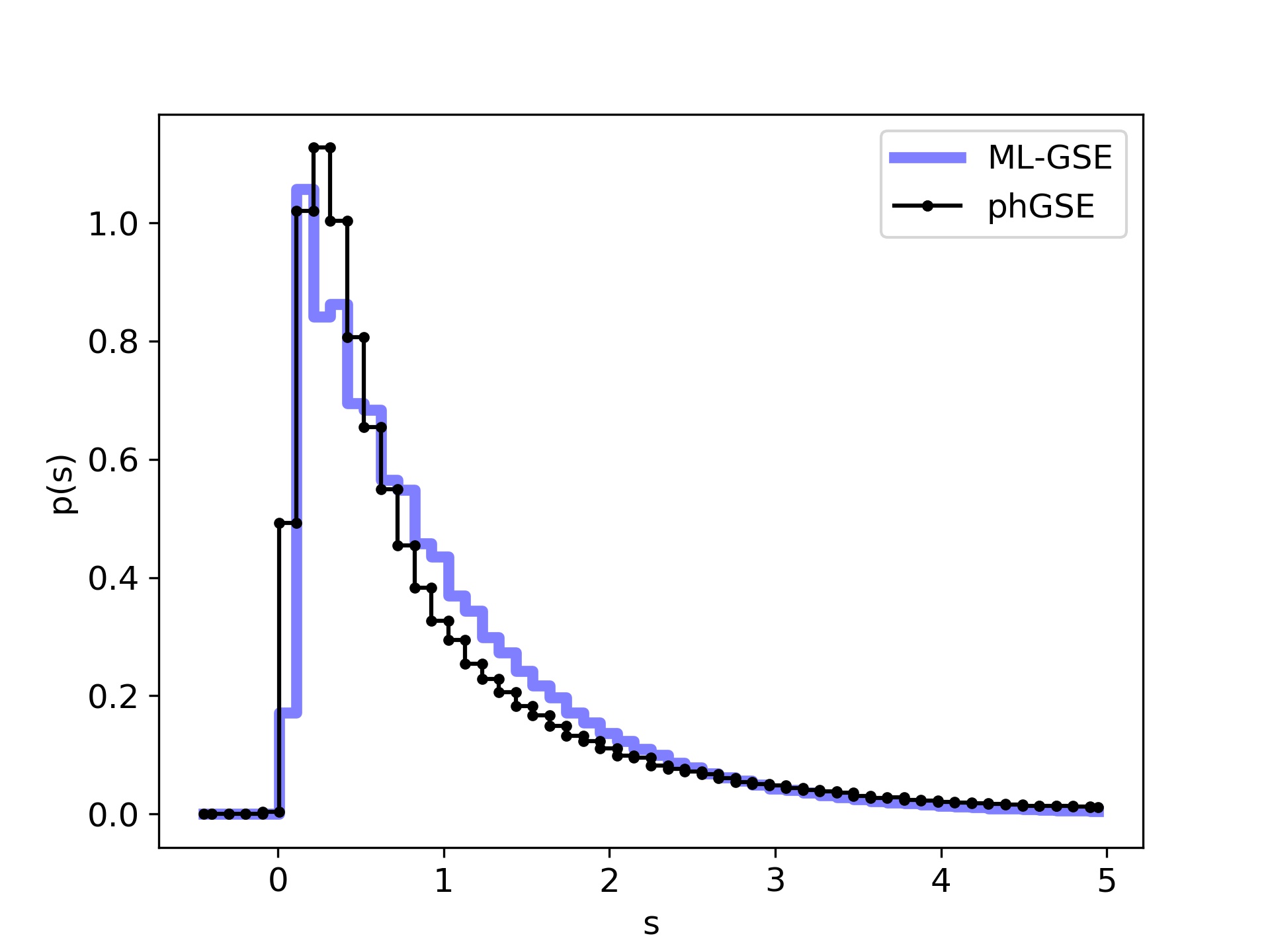}}
	\caption{\label{fig:sp:pHGSE} Black: spacing distribution of the real eigenvalues of the pHGSE ($\beta=4$) with $r=0.50$, with $N = 360$ and $M =90$; blue: spacing distribution of Hermitian GSE spectrum with only two levels, chosen at random, remaining. In both cases, a sample size of $2 \cdot 10^4$ was used.}
\end{figure}

Therefore, we may compare directly the spacing of the level spacing from the pseudo-Hermitian cases to that of the missing level cases. This was the procedure followed in  Figs. \ref{fig:sp:pHGOE}, \ref{fig:sp:pHGUE} and \ref{fig:sp:pHGSE}. In these three figures, the spacing distribution of the levels is compared with the spacing distributions of GOE, GUE and GSE spectra in which a fraction $f$ of levels is removed at random. The agreement obtained in the comparison show that the eigenvalues that remains in the real axis of the phGOE and phGUE ensembles indeed behave like levels of a randomly thinned ensemble, and agree well with the prediction from the number variance. The phGSE, however, does not follow the same behavior, although it still retains qualitative similarities to the minimal case of the missing levels model, the one in which all but two eigenvalues are removed. Notably, it still retains the repulsion behavior near the origin.

\subsection{Complex eigenvalues statistics}

Considering now the complex part of the spectra, we take as quantity of interest the probability of having, at the bulk of the region of the complex plane filled with the eigenvalues, an empty disk of radius $t.$ More relevant is the case when one of the eigenvalues is precisely located at the center of the empty disk. In this case, if $P(t)$ is the empty disk probability then $p(t)= -\frac{dP}{dt}$ is the probability of finding another eigenvalue at a distance $t$ of the center. Therefore, $p(t)$ is the spacing distribution and we have the relation  
\begin{equation}
	\left<t\right>=\int_{0}^{\infty} t\ p(t)\ dt=\int_{0}^{\infty} t\ \frac{d[1-P(t)]}{dt}\  dt=[1-P(t)]\ t\Big]_{0}^{\infty}+\int_{0}^{\infty}P(t)\ dt=\int_{0}^{\infty}P(t)\ dt
\end{equation}
that provides a practical way to extract the rescaled variable $s=t/\left<t\right>$ from the empty disk probability. 

We start by calculating the equivalent in two dimensions of the Poisson statistics as described in the appendix. The result obtained is shown in Fig. \ref{fig:Comulative} where the function $1-P(s)$ was fitted with the regularized incomplete gamma function  
\begin{equation}
	{\cal P}[(\kappa s)^2,\mu]=\frac{1}{\Gamma(\mu)}\int_{0}^{(\kappa s)^{2}}\exp(-t)\ t^{\mu -1}\ dt, 
	\label{G1}
\end{equation}
where 
\begin{equation*}
	\kappa =\Gamma\left(\frac{2\mu+1}{2}\right)/\Gamma\left(\mu\right).
\end{equation*}
The spacing distribution associated to this function is 
\begin{equation}
	p(s)=\frac{d{\cal P}[(\kappa s)^2,\mu]}{ds}=\frac{2\kappa^{2\mu}}{\Gamma(\mu)}e^{-\kappa^2 s^2}s^{2\mu -1}.
	\label{eq:spc}
\end{equation}
The fitting shown in the Fig. \ref{fig:Comulative} was obtained with $\mu=1$ . Therefore, uncorrelated points filling the ellipsis, repel each other as GOE levels, in one dimension, do. Of course, this result must be considered an effect of geometry.

\begin{figure}[ht]
	{\includegraphics*[width=0.49\textwidth]{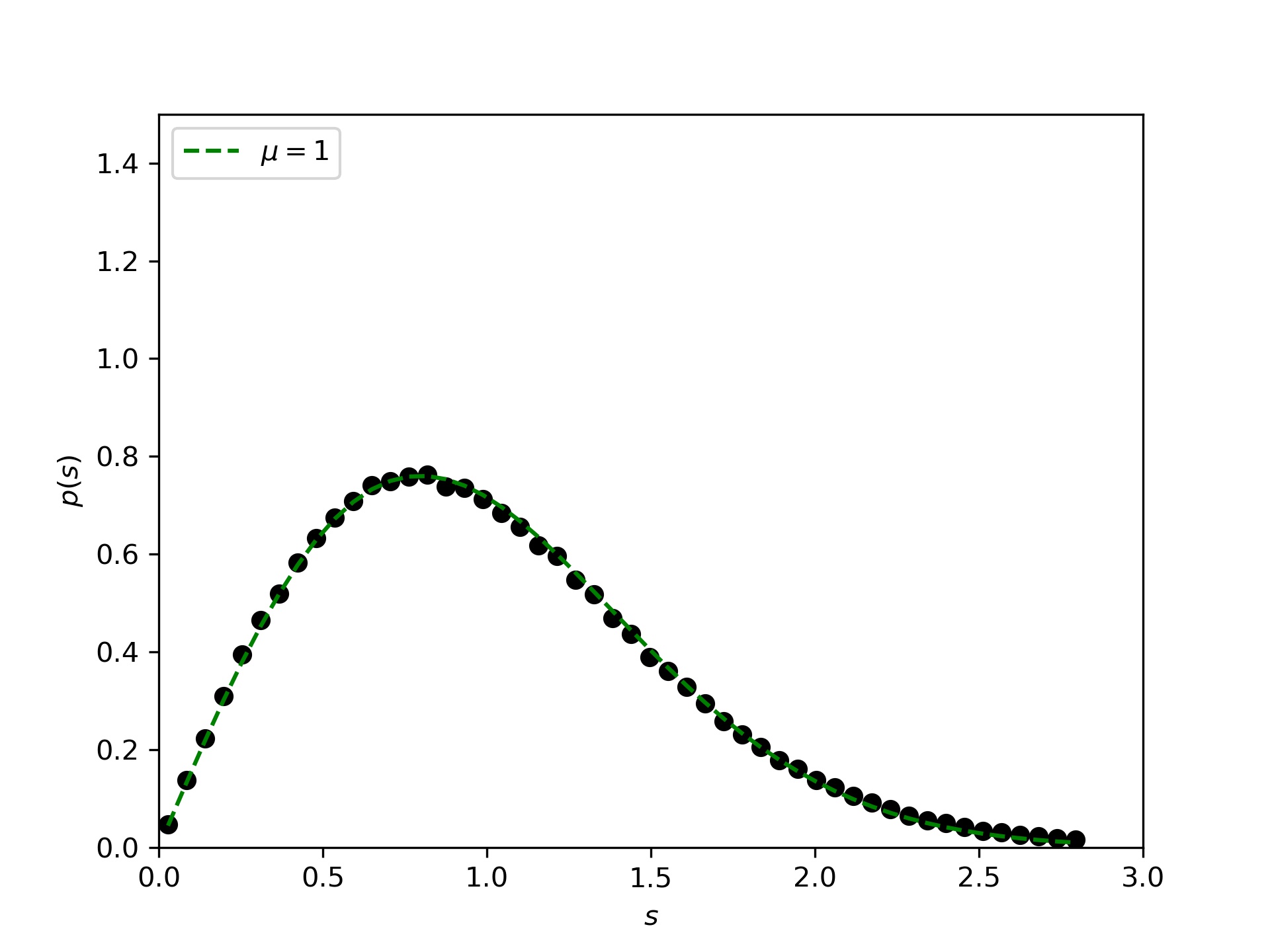}\includegraphics*[width=0.49\textwidth]{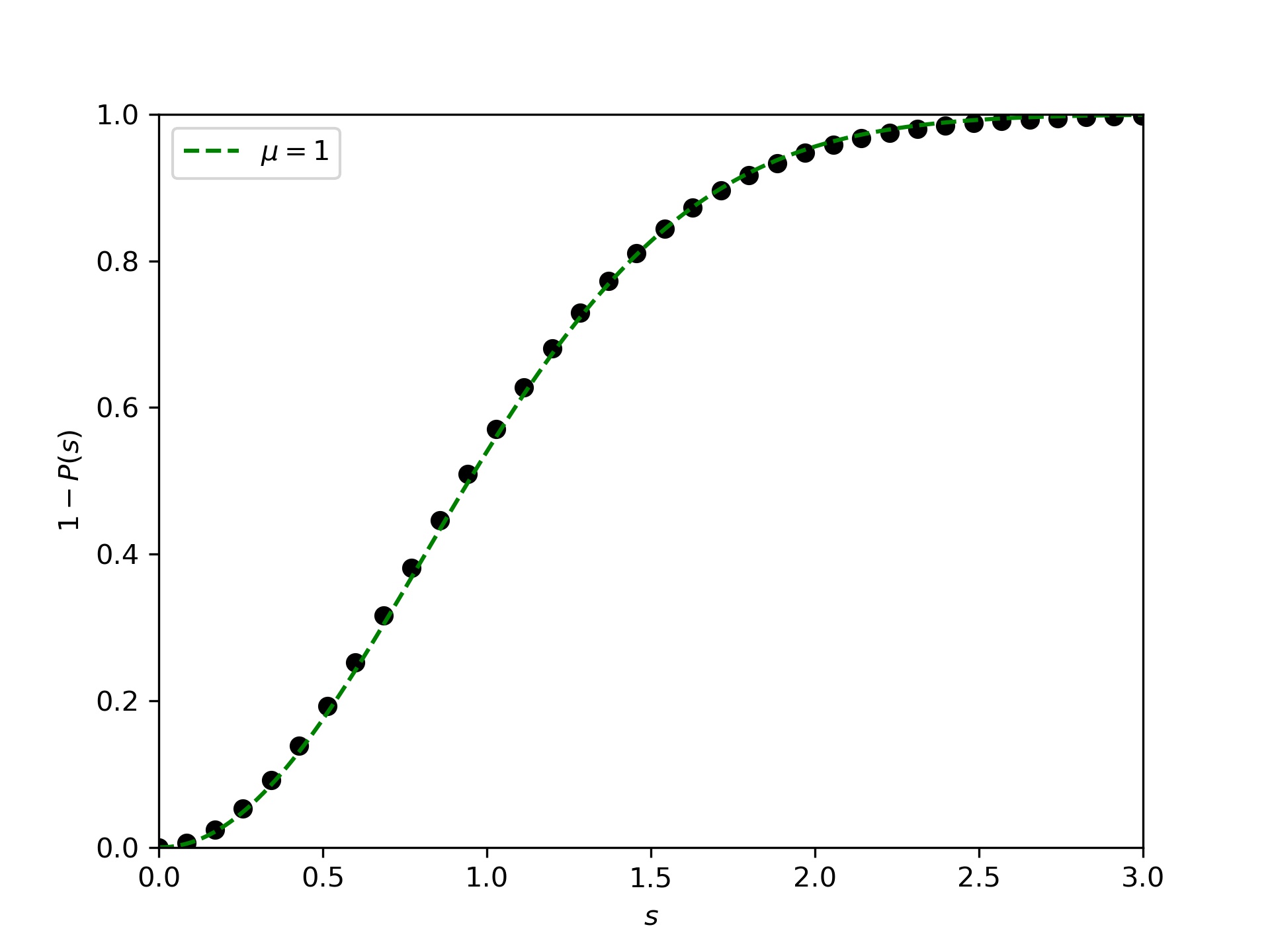}}
	\caption{\label{fig:Comulative} Black dots are the spacing (left) and cumulative (right) distributions calculated filling, with N=360 uncorrelated points, the ellipsis with $r=0.5$; the red lines are Eqs. \eqref{eq:spc} (left) and \eqref{G1} (right) with $\mu=1$.}
\end{figure}

\begin{figure}[ht]
	{\includegraphics*[width=0.49\textwidth]{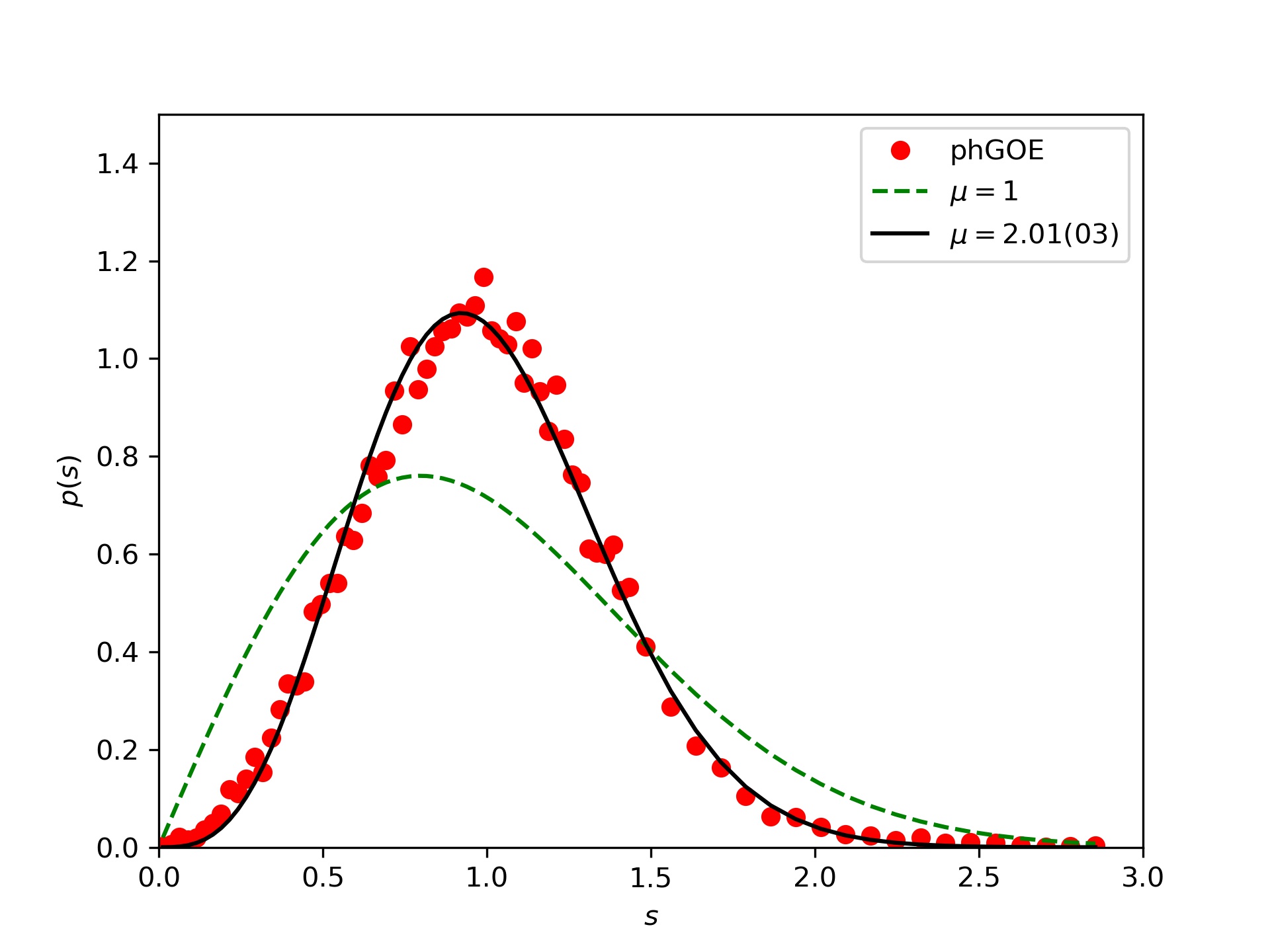}\includegraphics*[width=0.49\textwidth]{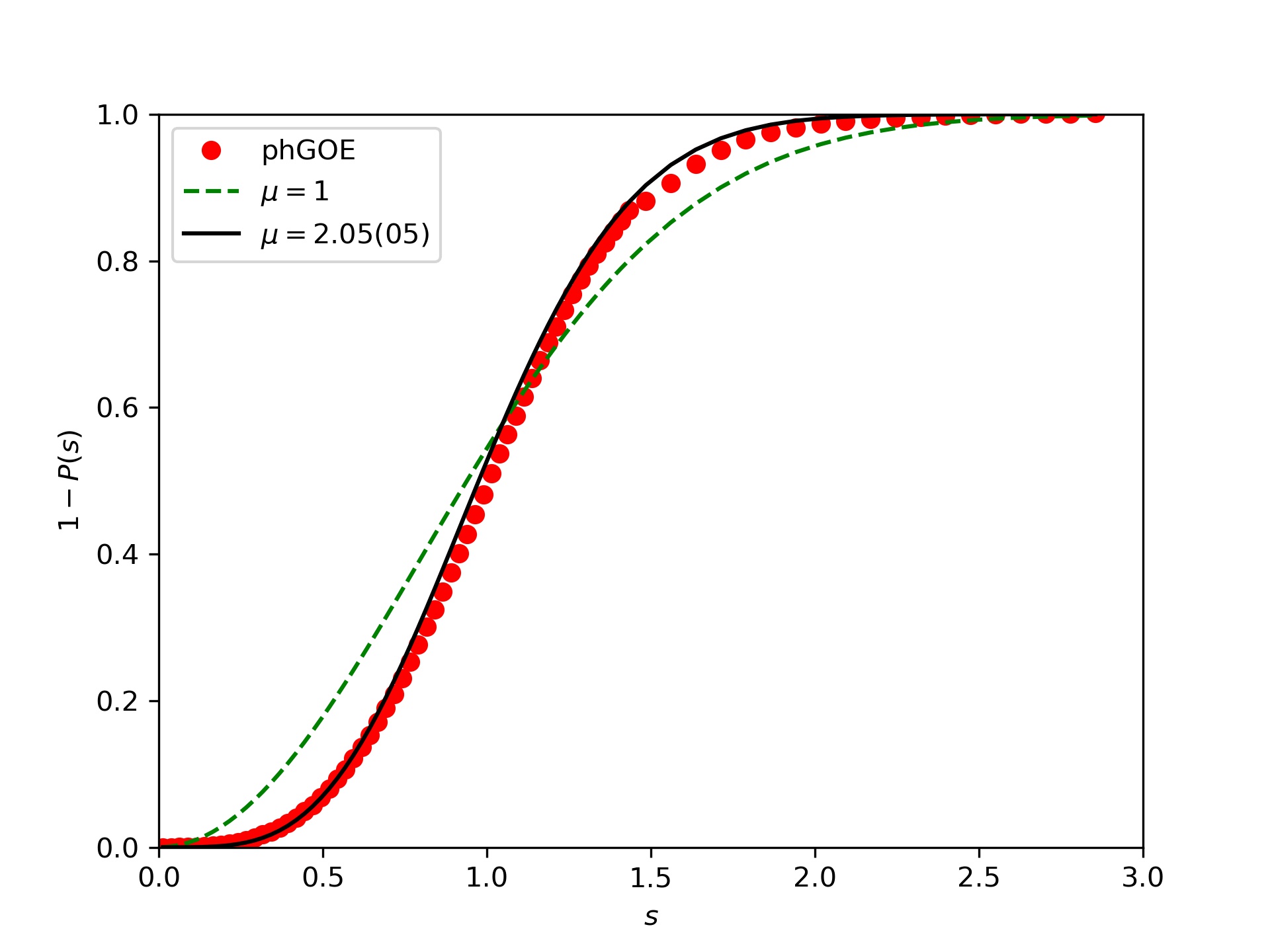}}
	\caption{\label{fig:complexspc:GOE} Red dots are spacing (left) and cumulative (right) distributions for the pHGOE, for $N = 360$, $M = 180$ and $r = 0.5$ (blue dots); the black lines are Eqs. \eqref{eq:spc} (left) and \eqref{G1} (right) fitted over the phGOE data, and also plotted are the same Eqs. with $\mu=1$ for the green dashed line.  Fit error was obtained from the Levenberg-Marquardt method.}
\end{figure}

\begin{figure}[ht]
	{\includegraphics*[width=0.49\textwidth]{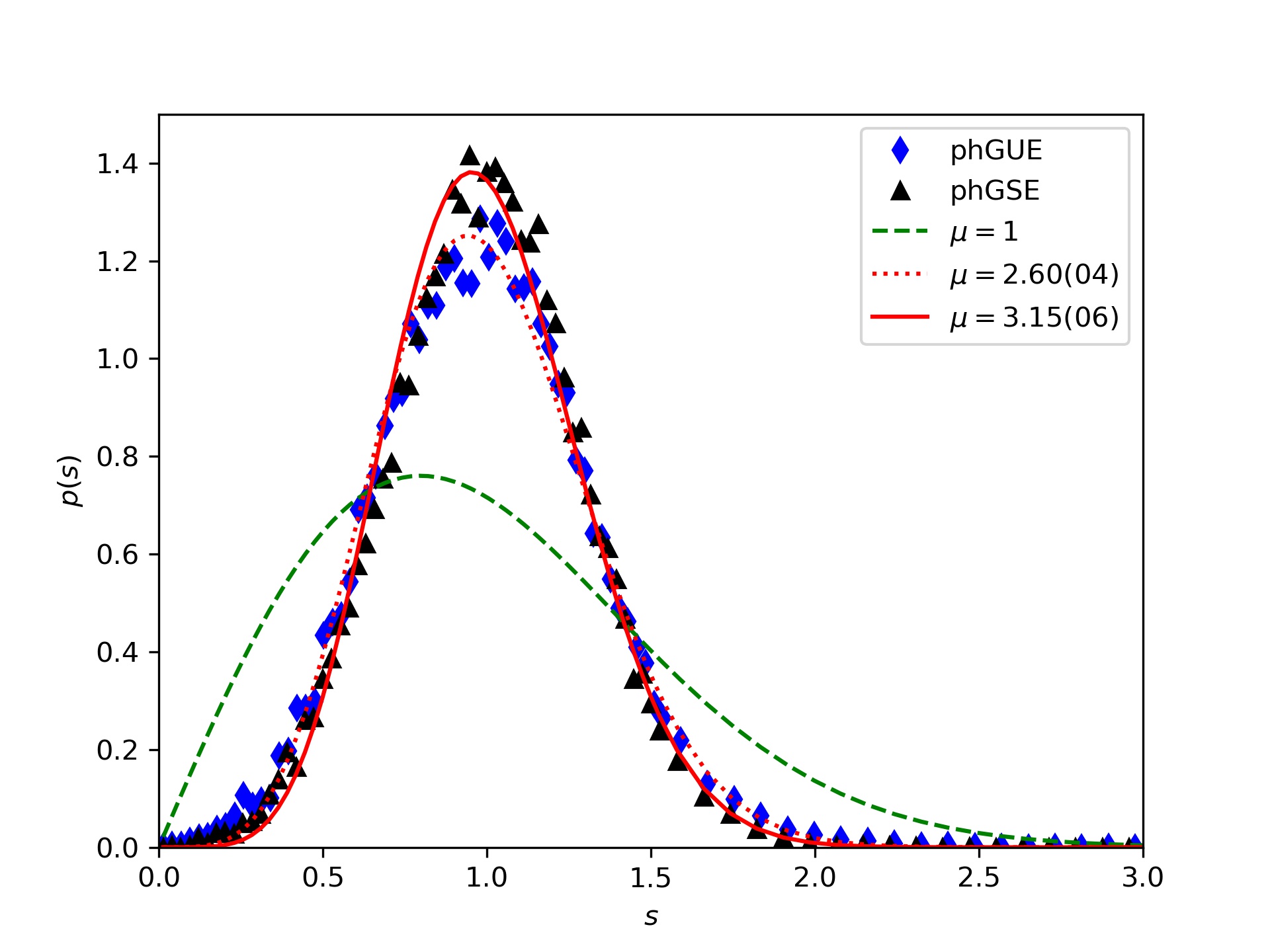}\includegraphics*[width=0.49\textwidth]{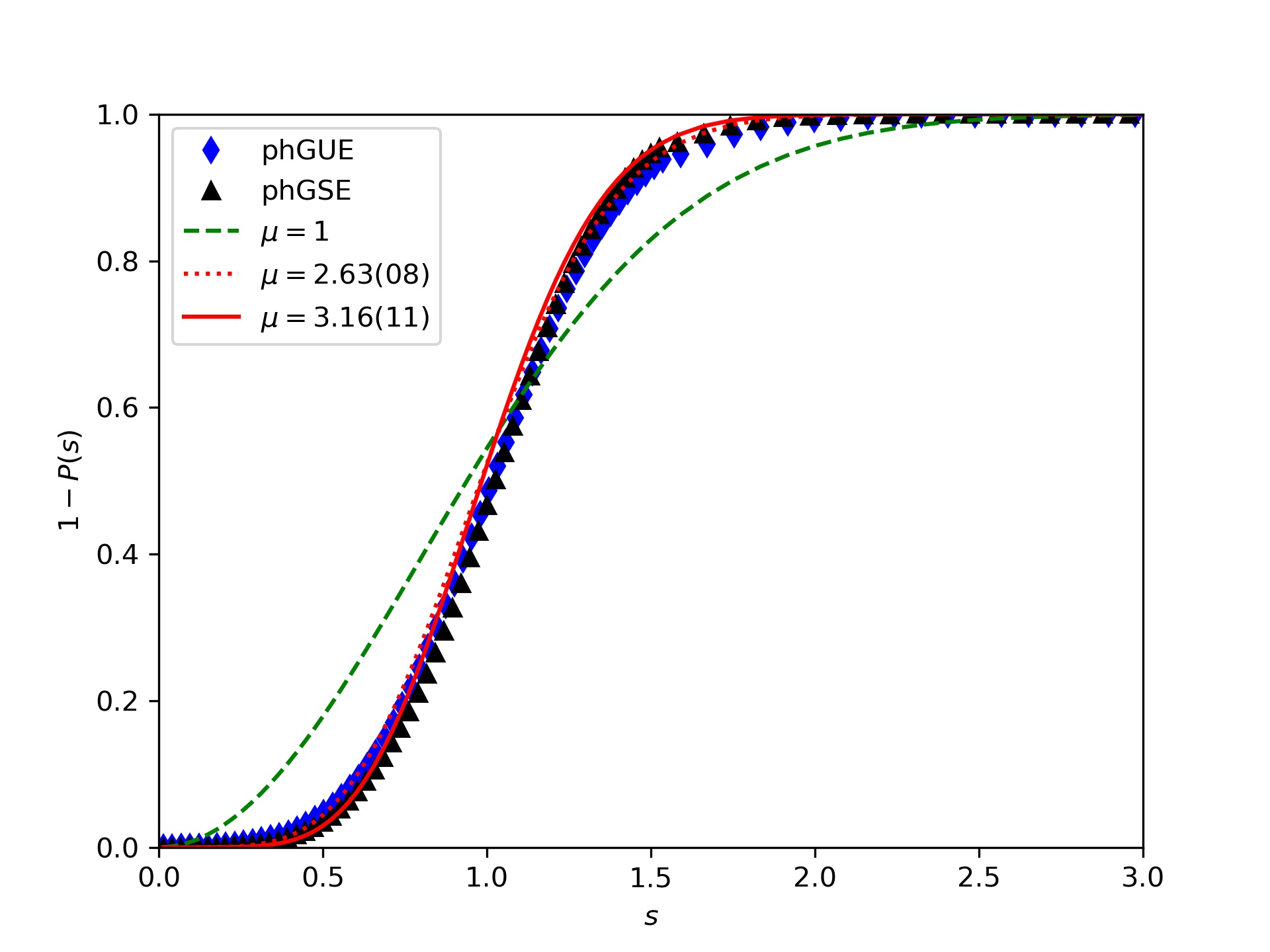}}
	\caption{\label{fig:complexspc:else}Black and red dots are spacing (left) and cumulative (right) distributions for the pHGUE, for  for $N = 360$, $M = 180$ and $r = 0.5$ (blue rhombi) and pHGSE, for the same parameters (black triangles); the dotted and solid red lines are Eq. \eqref{eq:spc} (left) and Eq. \eqref{G1} (right) fitted over the phGUE and phGSE data, respectively, and also plotted are the same Eqs. with $\mu=1$ for the green dashed line. Fit errors were obtained from the Levenberg-Marquardt method.}
\end{figure}

Considering, now, the complex eigenvalues of the pseudo-Hermitian ensemble, the results for the intermediate parameter $r = 0.5$ are shown in Figs. \ref{fig:complexspc:GOE} and \ref{fig:complexspc:else}. In Fig. \ref{fig:complexspc:GOE}, it is shown that for the pHGOE class, the result can be fitted with both Eqs. \eqref{eq:spc} and \eqref{G1}, yielding a value compatible with $\mu=2$ which implies in a cubic repulsion. In Fig. \ref{fig:complexspc:else}, it is shown that for the pHGUE and pHGSE classes, the results also can be fitted with the same Eqs. yielding values compatible with $\mu=2.5$ and $\mu=3.0$, which mean quartic and quintic repulsion for the phGUE and phGSE, respectively. This implies that our model follows the expected cubic repulsion only for the phGOE case, and the two remaining cases display higher-order repulsion. 

\begin{figure}[ht]
	{\includegraphics*[width=0.49\textwidth]{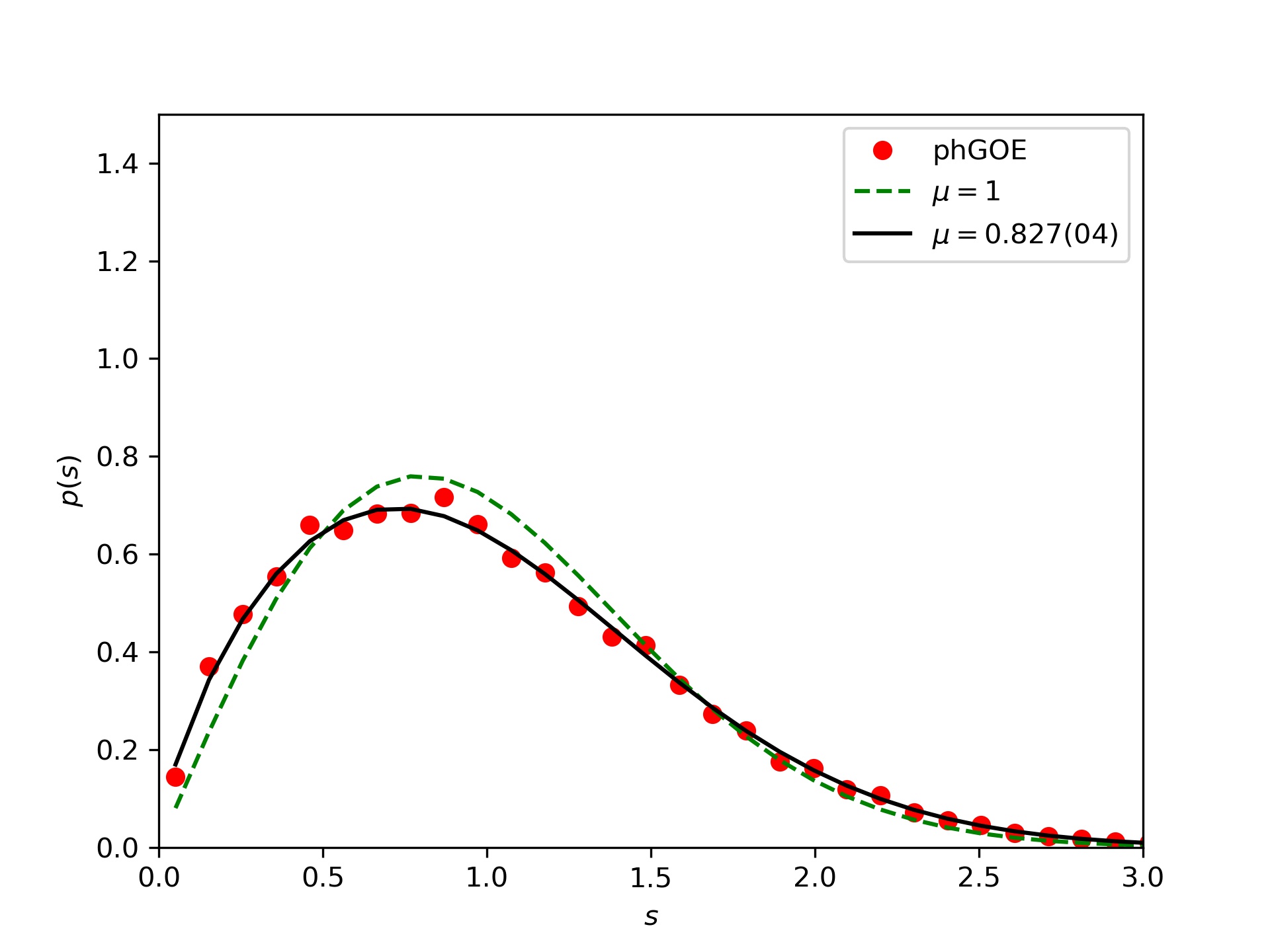}\includegraphics*[width=0.49\textwidth]{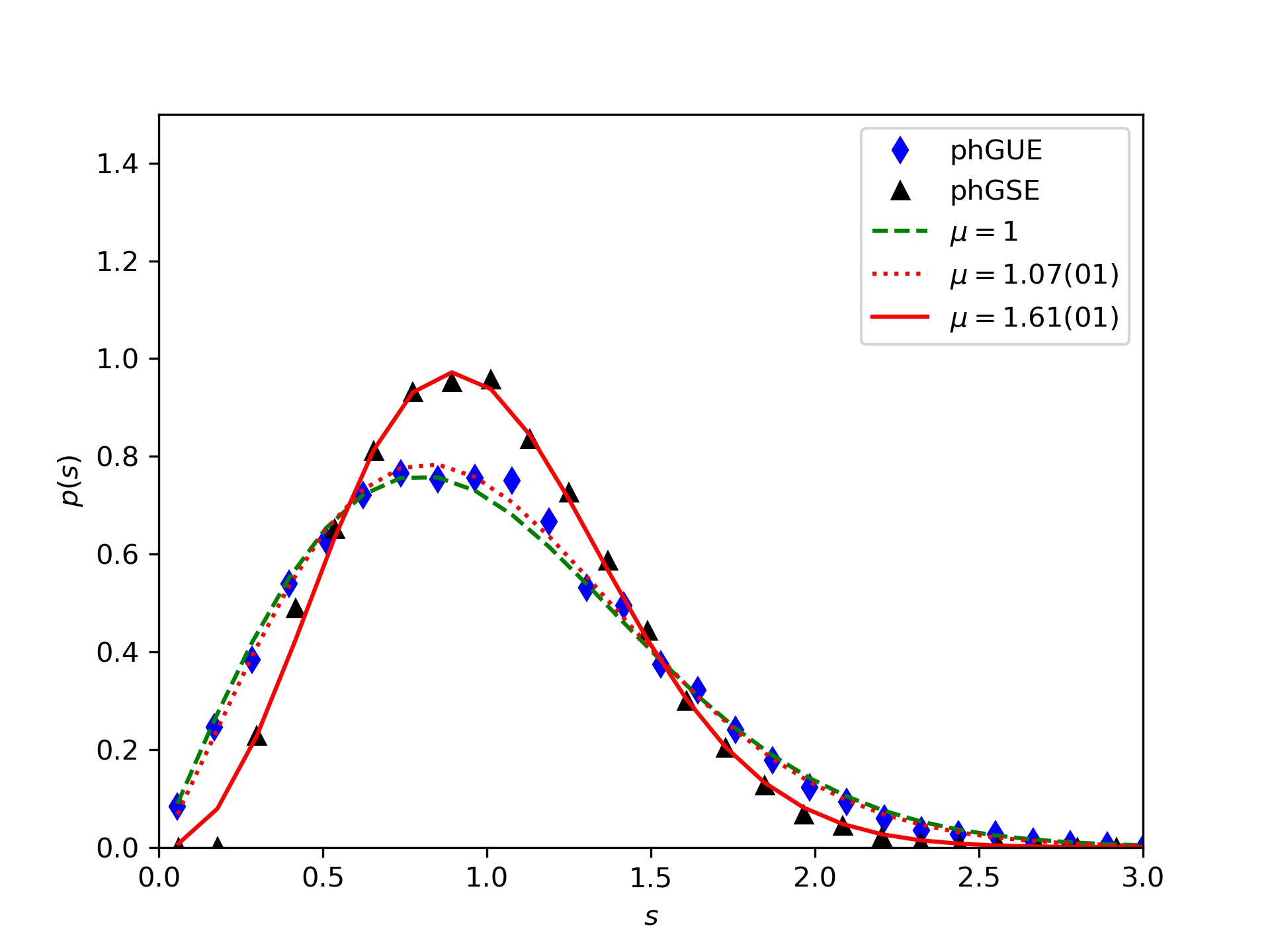}}
	\caption{\label{fig:complexspc:lowr}Black and red dots are spacing for the pHGOE (left) and phUE and phGSE (right), for  for $N = 360$, $M = 180$ and $r = 0.05$; the dotted and solid red lines are Eq. \eqref{eq:spc} (left) fitted over the data, and also plotted is the same Eq. with $\mu=1$ for the green dashed line. Fit errors were obtained from the Levenberg-Marquardt method.}
\end{figure}

\begin{figure}[ht]
	{\includegraphics*[width=0.49\textwidth]{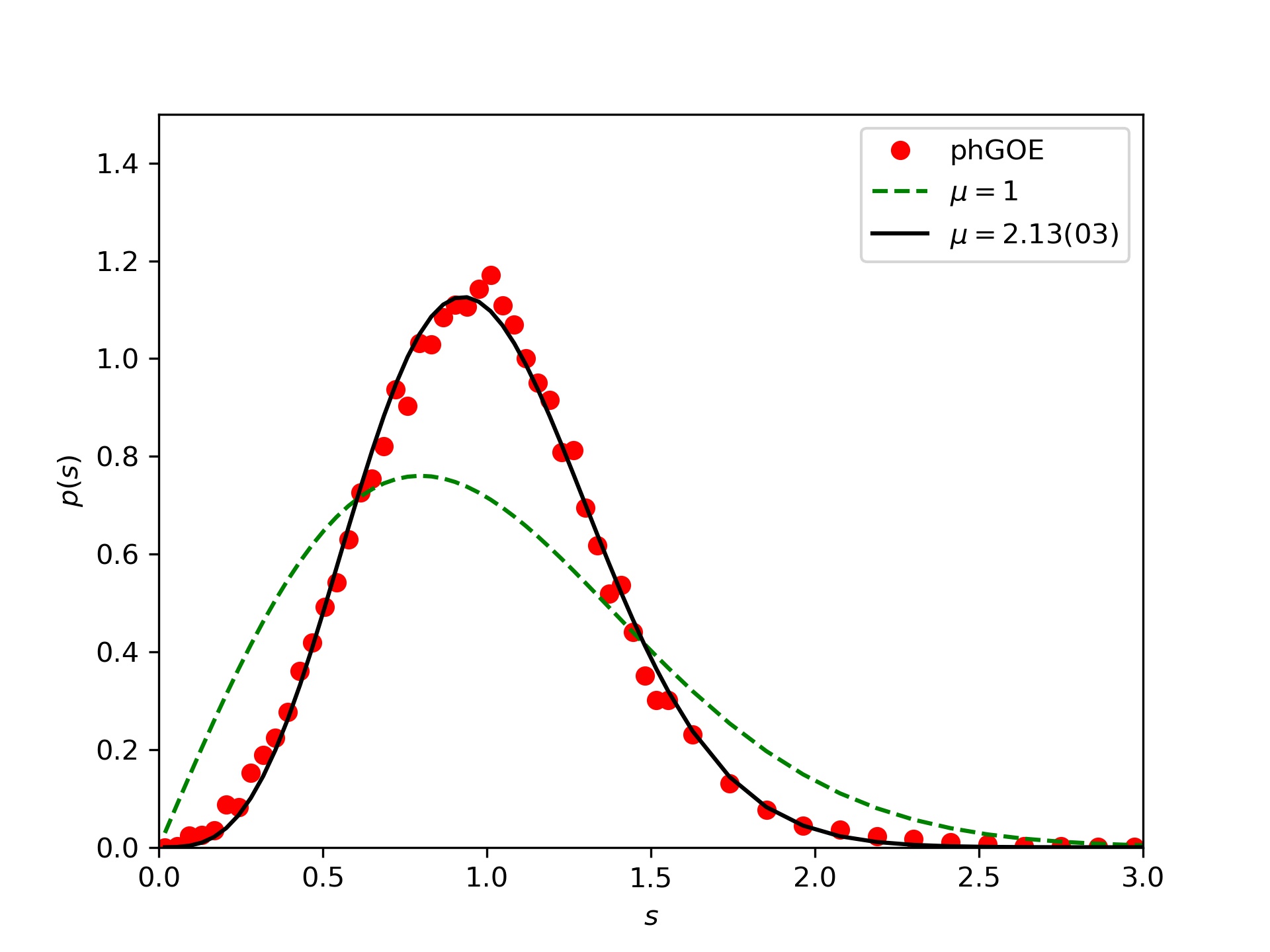}\includegraphics*[width=0.49\textwidth]{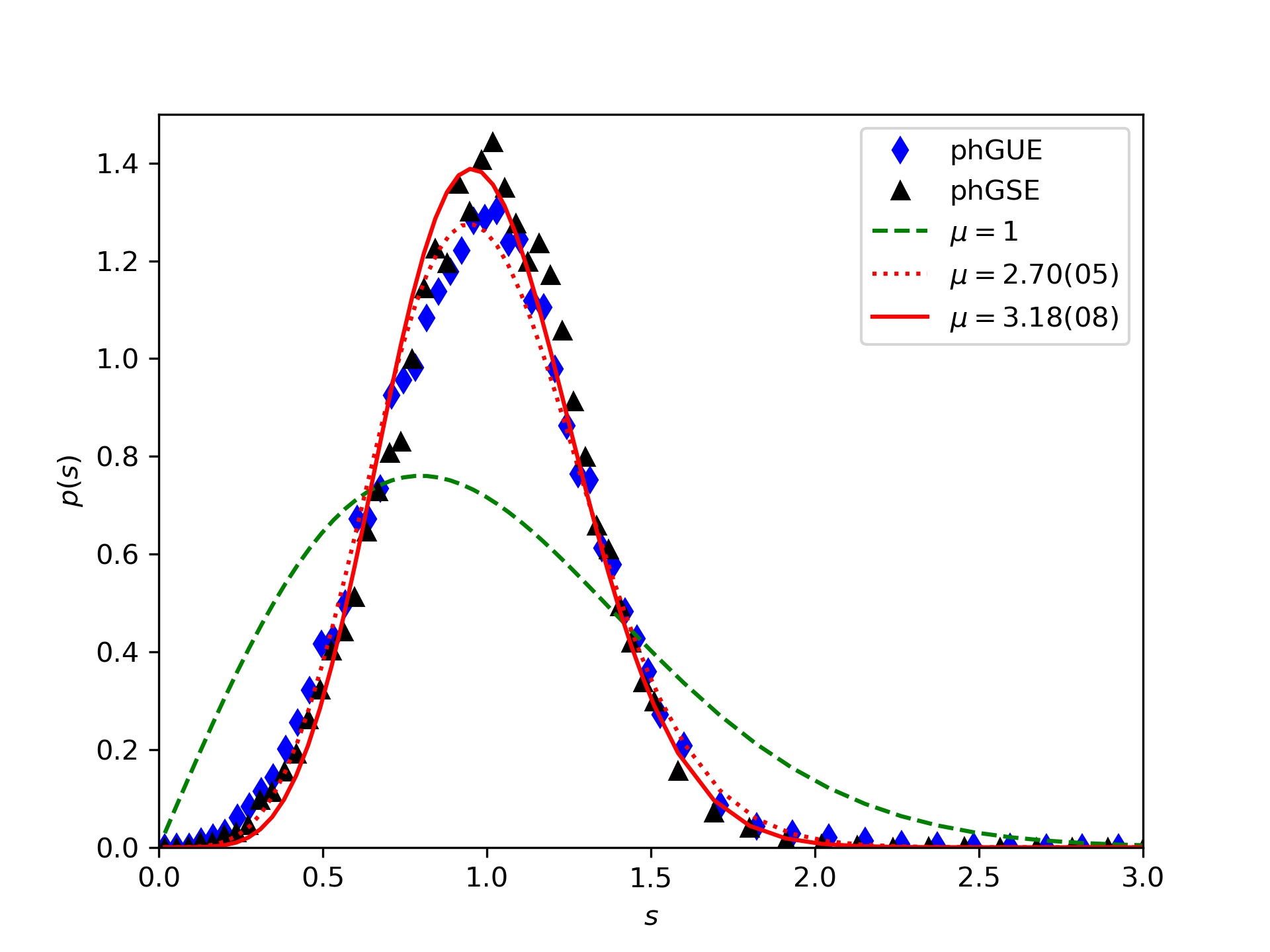}}
	\caption{\label{fig:complexspc:higr}Black and red dots are spacing for the pHGOE (left) and phUE and phGSE (right), for  for $N = 360$, $M = 180$ and $r = 2.0$; the dotted and solid red lines are Eq. \eqref{eq:spc} (left) fitted over the data, and also plotted is the same Eq. with $\mu=1$ for the green dashed line. Fit errors were obtained from the Levenberg-Marquardt method.}
\end{figure}

In Fig. \ref{fig:complexspc:lowr}, the same calculations are presented for $r = 0.05$, and it is seen that the behavior does not match that of the intermediary $r$. This is not surprising, as for low values of $r$ the spectra of these matrices are mostly real \cite{Marinello2016d}, and this represents a transitory regimen. This idea is corroborated by Fig. \ref{fig:complexspc:higr}, where the same plots are presented for the higher value of $r$, for which the behavior is very similar to that of the intermediate $r$.  

\section{Conclusion}

Let us first remark that any real matrix has eigenvalues that are real or complex conjugate. As a consequence, one should expect that a matrix $\eta$ must exist such that the matrix is connected to its adjoint by Eq. \eqref{11}. In fact, this is indeed the case, as a matrix $\eta$ can be constructed directly using the eigenvectors. The importance of the pseudo-Hermitian ensemble described in the second section is that, besides having elements that can be real, complex or quaternion, it has a fixed metric independent of the individual matrices.  

In a recent paper \cite{Marinello2018}, we have studied the spectral properties of the ensemble analyzing the average and the variance of its characteristic random polynomials. Here, we are approaching the eigenvalues statistical properties from the usual standard point of view calculating the spacing distributions of the real and the complex conjugate eigenvalues. One important result of the analysis is that, as  it was conjectured \cite{Bohigas2006}, the spectrum of the real eigenvalues behaves as if levels have been removed, at random, from the real axis. We also have observed the occurrence of a cubic repulsion between complex eigenvalues of the pHGOE class. This kind of repulsion has been reported for normal non-Hermitian matrices, that is matrices that commute with their adjoints, a property that the matrices of our ensemble do not have. For the remaining cases, however, the repulsion was found to be greater than cubic, an effect the cause of which remains an open question.

\section{Acknowledgments}

MPP acknowledges fruitful discussions with Rashid G. Nazmitdinov. GM was supported by grant 2019/00184-0 from the Brazilian agency FAPESP. MPP was supported by grant 307807/2017-7 from the Brazilian agency CNPq and is a member of the Brazilian National Institute for Science and Technology - Quantum Information (INCT-IQ). Additionally, we would to thank the anonymous referees for their suggestions, which have prompted us to expand our analysis in a very fruitful manner.

\appendix

\section{Poisson distribution}

In order to derive the Poisson distribution in a bidimensional space of a given shape one fills it, at random, with uncorrelated points. As a consequence, the probability of having a point inside the element of surface $dS$ is
\begin{equation}
\frac{dS}{S},
\end{equation}
where $S$ is the total area. For the ellipsis with axis defined in Ref. \cite{Marinello2016d}, 
elliptic coordinates $(u,v)$ are related to the Cartesian (x,y) as
\begin{equation}
x = \sqrt{\frac{N(1-r^2)}{1+r^2}}\cosh u\cos v 
\end{equation}
and
\begin{equation}
y = \sqrt{\frac{N(1-r^2)}{1+r^2}}\sinh u\sin v ,
\end{equation}
with the element of surface given by 
\begin{equation}
dS=\frac{N(1-r^2)}{1+r^2}\left(\cosh^2 u- \cos^2 v\right)dudv.
\end{equation}
Since the total area is $S=\pi ab$, we find that the above probability becomes 
\begin{equation}
P(u,v)du dv=\frac{\cosh^2 u- \cos^2 v}{\pi\sinh u_{0}\cosh u_{0}}du dv,
\end{equation}
where $u_0 =\mbox{atanh} (r^2)$. 

To generate a pair of values $(u,v)$, we start using the probability 
\begin{equation}
P(u_1)=\int_{0}^{2\pi}P(u_1,v) dv=\frac{2\cosh 2u_1} {\sinh 2u_{0}}
\end{equation}
of $u$ have a value $u=u_1$ disregarded of the value the other variable $v$ has. Then, 
once the value $u=u_1$ is obtained, the value of $v$ is extracted from the 
conditional probability 
\begin{equation}
P(v | u_1)=\frac{P(u_1,v)}{P(u_1)}=\frac{\cosh^2 u_1 - \cos^2 v} {\pi\cosh 2u_1} .
\end{equation}

\bibliography{refs}

\end{document}